\documentclass[man,11pt,floatsintext,a4paper]{apa7}

\usepackage{amsmath, amssymb}
\usepackage{graphicx}
\usepackage{booktabs}
\usepackage{hyperref} 
\usepackage{geometry}
\usepackage[dvipsnames]{xcolor}
\usepackage{xcolor}
\usepackage{tikz}
\usetikzlibrary{arrows.meta,positioning,fit,calc}
\usetikzlibrary{shapes.misc}
\usepackage{ulem}

\usepackage[natbibapa]{apacite} 
\usepackage{setspace}
\usepackage{verbatim}

\usepackage{float}         
\usepackage{threeparttable}

\title{The Force Concept Inventory Across Continents: Testing Q-Matrix Transferability and Cross-Cultural 
Differences in Mechanics Reasoning}
\shorttitle{The FCI Across Continents}
\authorsnames[{1,3},{2},{3,5},4]{Wade Naylor, Raheeq Tahir, Alan S. Cornell, Xuelan Qiu}
\authorsaffiliations{{School of Education, Australian Catholic University, Australia},
{Technolyceum (Ministry of Education School), Moscow Oblast, Russia},
{Department of Physics, University of Johannesburg, PO Box 524, Auckland Park 2006, South Africa},
{School of Education, City University Macau, Macau},
{Department of Physics, De La Salle University, 2401 Taft Avenue, Manila, 1004 Philippines}}

\abstract{
The Force Concept Inventory (FCI) is one of the most widely used research-based assessments in physics education, yet 
the assumption that its underlying cognitive structure is transferable across educational contexts remains largely untested. 
This study investigates the transferability of FCI Q-matrices using the Generalized Deterministic Inputs, Noisy “And” Gate (G-DINA) 
cognitive diagnostic applied to two large cohorts: students from the Learning About STEM Student Outcomes (LASSO) online system 
database in the United States (N = 4,750) and introductory physics students at the University of Johannesburg, 
South Africa (N = 1,016). 
Rather than treating the analysis as a local model calibration exercise, we frame the problem as one of cross-context 
cognitive invariance. Differential Item Functioning (DIF) analyses revealed substantial cross-cultural differences, with 14 of 30 
items exhibiting high DIF after controlling for latent skill mastery. These differences were concentrated in force dynamics and 
contact-force reasoning and remained invariant under alternative Q-matrix specifications. The findings suggest that observed 
differences reflect genuine variations in students’ conceptual reasoning rather than psychometric artifacts, highlighting the 
importance of validating Q-matrix structures before deploying cognitive diagnostic and adaptive assessments across diverse 
non-local educational settings.
}

\date{\today}

\begin{document}
\maketitle

\section{Introduction}

Improving student learning in science, technology, engineering, and mathematics (STEM) is an important goal in any context, 
be it in higher education (HE) or in a school based context. Aside from the development of curriculum, planning, and assessment 
for that particular subject, it also demands evidence-based approaches grounded in rigorous research. Within Discipline-Based 
Education Research (DBER), with Physics Education Research (PER) being a 
subset, academics have emphasized the need for tools that accurately measure conceptual understanding rather than rote 
memorization~\citep{NRC:2012, McDermott1999}. This need has led to the development and widespread adoption of 
Research-Based Assessments (RBAs), for example see \citet{affriyenni:2025} for an interesting discussion of RBAs' 
implementation. 

RBAs differ fundamentally from conventional assessments, where they are constructed through systematic research on student 
thinking and validated across diverse populations. They are further refined to ensure reliability and diagnostic power 
\citep{Adams:2006, Madsen2017_BestPractices}. In many cases, they are 
based on concept inventories (CIs) with multiple choice (MC) responses. One of the reasons for this being the time required to 
deploy and evaluate such CIs. 
The primary purpose of CIs is to uncover misconceptions and gaps in understanding, enabling instructors to tailor instruction and 
researchers to evaluate the effectiveness of pedagogical interventions~\citep{Singer2012}. By providing standardized benchmarks, 
RBAs also facilitate comparisons across courses, institutions, and instructional methods \citep{Henderson2011}.

One of the most influential RBAs in PER is the Force Concept Inventory (FCI), also known as the ``multi-million dollar 
CI'' \citep{Naylor:2024fci} (due to the extensive research hours and 100s of papers invested in the topic), developed by 
\citet{Hest:1992}.
The FCI targets students’ understanding of Newtonian mechanics, a compulsory facet of all 
undergraduate physics curriculum where persistent misconceptions often hinder learning. Unlike traditional problem-solving tests, the 
FCI probes conceptual reasoning, asking students to choose between scientifically accurate explanations and common intuitive 
alternatives. This design makes the FCI a powerful diagnostic tool for identifying misconceptions about force and motion 
\citep{Hest:1998, Hall:1985}. The impact of the FCI on physics education has been profound. Research using FCI data revealed that traditional lecture-based instruction often fails to produce significant conceptual gains, whereas interactive engagement strategies lead to dramatic improvements \citep{Hake:1998, Redish2003}. These findings have catalyzed widespread reforms in physics teaching, 
promoting active learning approaches that align with how students construct knowledge.

Traditionally, analyses of FCI data have relied on Classical Test Theory (CTT) measures such as normalized 
gains and total scores \citep{steyer_classical_2001}. More recently, Item Response Theory (IRT) has provided a stronger psychometric 
framework by modeling the relationship between latent ability and item responses~\citep{Rasch:1960,Birnbaum:1968}, enabling the 
development of computerized adaptive testing (CAT) systems that can substantially reduce testing time while maintaining measurement 
precision \citep[e.g., see][]{Yasuda:2021}. As such, these studies have shown how IRT-based adaptive assessment in 
STEM contexts can demonstrate personalized diagnostic testing and more efficient estimations of student 
proficiency in shorter times \citep{Yasuda:2021,Naylor:2024}. However, although IRT models are effective at estimating overall ability, 
they typically provide limited information regarding the specific concepts, skills, and misconceptions underlying such student 
responses.\footnote{Multidimensional IRT approaches \citep[e.g., see discussion in][]{Naylor:2024} do give more information than 
standard IRT models in terms of the skills in a given CI \citep[e.g., see][for the FCI ]{Wang:2010}. See the Methodology Section for
further details.}

Cognitive Diagnostic Models (CDMs) address this limitation by representing knowledge as a set of fine-grained attributes linked to 
assessment items through a Q-matrix \cite{delaTorre:2009}, allowing researchers to identify which cognitive components students have 
mastered and which remain problematic. In comparison with traditional IRT approaches, CDMs therefore offer a richer description of 
conceptual understanding and are particularly well suited to RBAs such as the FCI; for work with a more general 
set of mechanics cognitive diagnostics, see \citeauthor{le:2025}. Such CDMs can reveal that the structure of students’ misconceptions 
is often as important as estimating their overall performance. Furthermore, the integration of CDMs with adaptive testing frameworks 
has emerged as 
a promising direction for future STEM assessment, combining the efficiency of computerized adaptive testing 
(CAT) with detailed cognitive diagnostics \citep{le:2025}; also see recent work in \citep{le:2026}.

Diagnosing misconceptions in PER through formative assessment is a pivotal practice \citep{Hest:1998}. 
According to \cite{allen_ross:2017} misconceptions usually involve the application of inappropriate rules or generalizations to 
solve a question. For example, many students believe that if a feather and an apple are dropped, the apple will reach the ground first. 
This belief reflects the common misconception that ``heavy objects fall faster than lighter objects'' in terms of forces and Newton’s 
second law. When CDMs and formative assessment are used together to diagnose misconceptions, individual profiles can indicate which misconceptions 
students possess and which they do not. In this regard, we need accurate, efficient, and personalized profiles; and importantly, 
profiles that can investigate questions of equity \citep{buncher:2025}. 
 
In this work, we will take an approach to using FCI data suitable for 
focusing on the use of cognitive diagnostics where skills/attributes will be used for generalized CDM 
models \citep{delaTorre:2009}. 
This will be applied to PER by considering the skills in the FCI for a large cohort of physics students at the 
University of Johannesburg (UJ; 2020--2024, $N=1,016$) and compare to the Learning About STEM Student Outcomes (LASSO) 
\citep{lasso:nodate,nissen_providing_2022} data. Fundamental to any type of CDM model is the ''question-attribute 
matrix'' (Q-matrix); for MC items, it maps attributes to specific skills, so that even an 
incorrect answer may indicate that a certain skill has been attained. This fine grained approach is one of the 
reasons why we seek to use CDMs, 
see the Methodology Section.

In summary, RBAs, and the FCI in particular, are indispensable for advancing evidence-based teaching in entry-level physics. They serve as both a 
mirror, reflecting students’ current understanding, and a compass, guiding educators toward more effective instructional practices. 
Such approaches exemplify the core mission of PER and DBER; leveraging research to transform  education and improve learning outcomes in diverse student 
contexts.

\section{Research Questions}
 
Based on the previous discussion and the study we have undertaken at the Department of Physics, UJ, we ask the following two 
questions:
 \begin{enumerate}
 \item  \it Can a FCI Q-matrix developed from United States cohorts be validly transported to a South African higher 
 education context?
 \item \it When comparing differential item functioning (DIF) between US and South African cohorts are there any 
 significant effect-sizes for items in the FCI?
 \end{enumerate}

In order to answer these questions, in the following sections we review CDMs in physics education, 
with particular emphasis on the FCI using Generalized 
Deterministic Inputs, Noisy “And” Gate (G-DINA) framework \citep{delaTorre:2009}.
We introduce the theoretical foundations of Q-matrix construction, attribute specification, and DIF within 
cognitive diagnostics, including a novel way to determine effect sizes from Cohen's $h$ \citep{cohen_statistical_2013}, 
see the Methods section.  
Next, we describe the datasets analyzed in this study, including the United States LASSO and the 
UJ cohort, together with procedures used for data cleaning, model calibration, 
and Q-matrix validation, see the Data Sources section in the Results. 
 
In the Results section 
we subsequently examine the transferability of the FCI Q-matrix across these distinct educational contexts by 
comparing item-level and attribute-level model behavior. This is followed by a detailed  DIF analysis to identify 
items that exhibit measurement non-invariance after controlling for latent skill mastery. We then investigate the 
implications of these differences for understanding cross-cultural variations in conceptual mechanics reasoning (using the 
FCI) and evaluate the robustness of our findings through Q-matrix sensitivity analyses. 
 
Finally, in the Conclusion 
we discuss the implications of these results for the use of CDMs in PER, the development of adaptive assessments in general, 
and future studies examining equity, fairness, and measurement invariance across diverse student populations.

\section{Methodology}
\label{sec:meth}


\begin{figure}[ht]
\begin{minipage}{\linewidth}
\centering 
\caption[]{
Analyzing the FCI with a cognitive diagnostic model (CDM)\footnote{Each examinee is represented by a categorical 
latent-attribute profile $\boldsymbol{\alpha}_i=(\alpha_{i1},\ldots,\alpha_{iK})$, where $\alpha_{ik}=1$ denotes 
mastery and $\alpha_{ik}=0$ denotes non-mastery for $K$ skills. The Q-matrix specifies the attributes required 
by each item: $q_{jk}=1$ if item $I_j$ requires attribute $k$, and $q_{jk}=0$ otherwise.}
}
\label{fig:cdm} 
\scalebox{0.8}{
\begin{tikzpicture}
[ >=Stealth, attribute/.style={ circle, draw, minimum size=1.65cm, align=center, font=\large }, profile/.style={ rounded rectangle, draw, 
minimum width=5.0cm, 
minimum height=1.0cm, align=center, font=\large }, qmatrix/.style={ rounded rectangle, draw, thick, minimum width=5.8cm, minimum height=1.25cm, 
align=center, font=\large }, 
item/.style={ rectangle, draw, minimum width=1.15cm, minimum height=1.0cm, align=center, font=\large }, model/.style={ rounded rectangle, draw, 
dashed, minimum width=6.0cm, minimum height=0.9cm, align=center, font=\small }, every edge/.style={ draw, ->, thick } 
] 
\node[profile] (profile) at (5.25,2.2) { Latent attribute profile: $\boldsymbol{\alpha}_i =(\alpha_{i1},\alpha_{i2},\ldots,\alpha_{iK})$ }; 
------------------------------------------------- 
\node[attribute] (K1) at (0.5,0) { $K_1$\\[1mm] $\alpha_{i1}\in\{0,1\}$ }; 
\node at (2.2,0) {\Large $\cdots$}; 
\node[attribute] (K2) at (3.8,0) { $K_2$\\[1mm] $\alpha_{i2}\in\{0,1\}$ }; 
\node at (5.5,0) {\Large $\cdots$}; 
\node[attribute] (K3) at (7.1,0) { $K_3$\\[1mm] $\alpha_{i3}\in\{0,1\}$ }; 
\node at (8.8,0) {\Large $\cdots$}; 
\node[attribute] (Kn) at (10.5,0) { $K_K$\\[1mm] 
$\alpha_{iK}\in\{0,1\}$ }; 
\draw[->,thick] (profile.south west) to[out=-120,in=90] (K1.north); 
\draw[->,thick] ($(profile.south)+(-1.1,0)$) to[out=-100,in=90] (K2.north); 
\draw[->,thick] ($(profile.south)+(1.1,0)$) to[out=-80,in=90] (K3.north); 
\draw[->,thick] (profile.south east) to[out=-60,in=90] (Kn.north); 
\node[qmatrix] (Q) at (5.25,-2.25) { $\mathbf{Q}=(q_{jk})$, \qquad $q_{jk}\in\{0,1\}$ \\[1mm] 
\footnotesize $q_{jk}=1$: item $j$ requires attribute $k$ }; 
\draw[->,thick] (K1) -- (Q.north west); 
\draw[->,thick] (K2) -- ($(Q.north)+(-1.1,0)$); 
\draw[->,thick] (K3) -- ($(Q.north)+(1.1,0)$); 
\draw[->,thick] (Kn) -- (Q.north east); 
\node[item] (I1) at (0.7,-4.6) {$I_1$}; 
\node[item] (I2) at (2.5,-4.6) {$I_2$}; 
\node at (3.75,-4.6) {\Large $\cdots$}; 
\node[item] (Ij) at (5.25,-4.6) {$I_j$}; 
\node at (6.85,-4.6) {\Large $\cdots$}; 
\node[item] (Im1) at (8.0,-4.6) {$I_{J-1}$}; 
\node[item] (Im) at (10.1,-4.6) {$I_J$}; 
\draw[->,thick] (Q.south west) to[out=-120,in=90] (I1.north); 
\draw[->,thick] ($(Q.south)+(-1.8,0)$) to[out=-105,in=90] (I2.north); 
\draw[->,thick] (Q.south) -- (Ij.north); 
\draw[->,thick] ($(Q.south)+(1.8,0)$) to[out=-75,in=90] (Im1.north); 
\draw[->,thick] (Q.south east) to[out=-60,in=90] (Im.north); 

\end{tikzpicture}
}
\end{minipage}
\end{figure}

 Beyond classical test theory \cite{steyer_classical_2001}, IRT has been well used as a way to find latent 
attributes that can be assigned to students and also where MC items can be assigned an item difficulty \citep{Naylor:2024}. 
IRT has both a unidimensional (UIRT) and multidimensional (MIRT) versions, where IRT in 
general allows researchers to check the suitability of each test item by quantifying item difficulty and student 
ability for each question item \citep[e.g., see][]{Naylor:2024}. The MIRT version allows each CI to be broken down into 
sub-concepts such as the six conceptual parts typically assumed for the FCI \citep{Chrysostomou:2024, Naylor:2024}, 
where it also allows for correlations and covariances between each sub-concept.

CDMs extend upon IRT models by not just looking at the correct/incorrect responses (dichotomous choices) in a 
MC test, but also considering knowledge attributes that are categorical in nature; see Figure \ref{fig:cdm} for a graphical 
representation. Such CDMs can be used to understand the skills, cognitive processes, and  problem-solving 
strategies involved in an assessment \citep{delaTorre:2009}. 

In this work we use a set of deterministic inputs, noisy ``And'' gate (DINA) model \citep{delaTorre:2009} that 
is able  to diagnose in an assessment item the skills 
and attributes used in a given correct response to an MC item. An example of the types of skills and 
misconceptions that might arise in the FCI and that have been used in CDMs already in the context 
of PER \citep{le:2025} is shown in Table \ref{table:skills}. It is worth mentioning that while the DINA model assumes a 
conjunctive structure, where all skills are assumed needed to successfully complete a question, 
we employ the G-DINA \citep{delaTorre:2011} framework which relaxes this assumption as we discuss next. 

\begin{table}[ht]
\centering
\caption{\small \it A list of the skills, $k_n$, used for the Q-matrix for the FCI, adapted from \protect\citet{le:2025}
        }
\begin{tabular}{p{0.25\linewidth} p{0.05\linewidth}   p{0.6\linewidth}}
\hline 
Skill & $k_n$ & Item \\ 
\hline
Apply Vectors & $k_1$    & Requires manipulating vectors in more than one dimension or has a change in sign for 
a 1-D vector quantity \\ 
Conceptual Relationships  & $k_2$ & Requires students to identify a relationship between variables and/or the 
situations in which those 
relationships apply \\ 
Visualizations & $k_3$ & Requires extracting information from or creating formal visualizations such as xy plots, 
bar plots, or line graphs \\ 
\hline
\end{tabular} 
\label{table:skills}
\end{table}

\subsection{DINA/G-DINA Cognitive Diagnostic Models}
\label{sec:CDM}

\noindent The DINA model \citep{delaTorre:2009} is a conjunctive CDM assuming that examinees must 
master all required attributes to answer an item correctly for a given set of skills $k_n$. The following definitions 
will be useful for explaining how these models work. There is a latent attribute profile for each examinee 
\( i \) that has a binary attribute vector (also see Figure \ref{fig:cdm}):
\begin{equation}
\boldsymbol{\alpha}_i = (\alpha_{i1}, \alpha_{i2}, \dots, \alpha_{iK}), \quad \alpha_{ik} \in \{0,1\},
\end{equation}
where the Q-matrix \( Q = (q_{jk}) \) specifies which attributes each item \( j \) requires:
\[
q_{jk} = \begin{cases}
1 & \text{if item } j \text{ requires attribute } k, \\
0 & \text{otherwise.}
\end{cases}
\]
Then for dichotomous responses, the probability that a response will be correct under the DINA model is given by:
\begin{equation}
P(Y_{ij}=1|\eta_{ij}) = (1-s_j)^{\eta_{ij}} g_j^{1-\eta_{ij}},
\end{equation}
where $s_j$ is the slip parameter (the probability of error when fully skilled) and $g_j$ is the guess parameter 
(the probability of a correct response when missing at least one skill). The ideal response indicator is defined by:
\begin{equation}
\label{eq:eta}
\eta_{ij} = \prod_{k=1}^K \alpha_{ik}^{q_{jk}},
\end{equation}
so that \( \eta_{ij}=1 \) if the examinee has all required attributes, and \( 0 \) otherwise.

\par To address the restrictive non-compensatory assumptions of the standard DINA model utilized in prior FCI diagnostic research 
\citep{le:2025}, 
this study employs the Generalized DINA (G-DINA) framework \citep{delaTorre:2011}. The G-DINA model relaxes deterministic restrictions 
by allowing for main \& interaction effects of mastery across required attributes.

For an item $j$ requiring a reduced attribute vector $\boldsymbol{\alpha}_{j}^* = (\alpha_1, \dots, \alpha_{K_j}^*)$, the success 
probability for an examinee in attribute mastery class $\boldsymbol{\alpha}_{j}^*$ is formulated using the identity link function as:
\begin{equation}
P(\boldsymbol{\alpha}_{j}^*) = \delta_{j0} + \sum_{k=1}^{K_j^*} \delta_{jk} \alpha_{k} + \sum_{k'=k+1}^{K_j^*} 
\sum_{k=1}^{K_j^*-1} \delta_{jkk'} \alpha_{k} \alpha_{k'} + \dots + \delta_{j12\dots K_j^*} \prod_{k=1}^{K_j^*} \alpha_{k},
\label{eq:gdina}
\end{equation}
where $\delta_{j0}$ is the baseline probability (guessing effect), $\delta_{jk}$ is the main effect of attribute $k$, 
$\delta_{jkk'}$ is the interaction effect between attributes $k$ and $k'$, and $\delta_{j12\dots K_j^*}$ is the highest-order 
interaction effect across all $K_j^*$ required attributes.

\subsubsection{Differential Item Functioning (DIF) in G-DINA}
\label{sec:gdina_dif}

\noindent Under the G-DINA framework, Differential Item Functioning (DIF) occurs when item parameters differ significantly across focal 
and reference groups (e.g., UJ vs. LASSO) for examinees with identical attribute mastery profiles. Formally, item $j$ exhibits 
DIF if:
\begin{equation}
P_g(\boldsymbol{\alpha}_{j}^*) \neq P_{g'}(\boldsymbol{\alpha}_{j}^*) \quad \text{for at least one } \boldsymbol{\alpha}_{j}^*,
\end{equation}
where $g$ and $g'$ denote distinct subgroup cohorts. Item parameter shifts across cohorts are evaluated using the item-level 
Wald test within the G-DINA framework \citep{hou2014differential}.

\begin{figure}
\begin{minipage}{\linewidth}
\centering
\caption[]{Conceptual illustration of DIF in a CDM\footnote{Groups A and B share the same attribute structure and Q-matrix, 
but Item 3 exhibits different item parameters across groups, indicating potential DIF after controlling for attribute mastery. 
For simplicity only three items $I_1,I_2,I_3\dots$ are shown for a general set of $I_n$ terms.}
}
    \label{fig:DIFCDM}
\begin{tikzpicture}[ skill/.style={circle,draw,minimum size=1cm}, item/.style=
{rectangle,draw,minimum width=1cm,minimum height=.8cm}, group/.style={rounded 
corners,draw,thick}, >=Stealth ] 
\node[skill] (A1) at (-2,1.5) {$K_1$}; 
\node[skill] (A2) at (0,1.5) {$K_2$}; 
\node[skill] (A3) at (2,1.5) {$K_3$}; 
\node[item] (I1A) at (-2,-1) {$I_1$}; 
\node[item] (I2A) at (0,-1) {$I_2$}; 
\node[item,fill=gray!20] (I3A) at (2,-1) {$I_3$}; 
\draw[->] (A1)--(I1A); \draw[->] (A2)--(I1A); 
\draw[->] (A2)--(I2A); \draw[->] (A3)--(I2A); 
\draw[->] (A1)--(I3A); \draw[->] (A3)--(I3A); 
\node[font=\small] at (0,-2.2) {Group A}; 

\node[skill] (B1) at (7,1.5) {$K_1$}; 
\node[skill] (B2) at (9,1.5) {$K_2$}; 
\node[skill] (B3) at (11,1.5) {$K_3$}; 
\node[item] (I1B) at (7,-1) {$I_1$}; 
\node[item] (I2B) at (9,-1) {$I_2$}; 
\node[item,fill=red!20] (I3B) at (11,-1) {$I_3$};
\draw[->] (B1)--(I1B); \draw[->] (B2)--(I1B); 
\draw[->] (B2)--(I2B); \draw[->] (B3)--(I2B); 
\draw[->] (B1)--(I3B); \draw[->] (B3)--(I3B); 
\node[font=\small] at (9,-2.2) {Group B}; 
\draw[very thick,red,->] (I3A) to[bend left=20] node[midway,above] {DIF} (I3B); 

\end{tikzpicture}

    \end{minipage}
\end{figure}

\subsection{Q-matrix Transferability and DIF in CDMs}

\noindent In this article we frame the analysis of the FCI as a problem of {\it Q-matrix transferability} across contexts 
rather than a local application of a CDM. Specifically, we investigate whether a Q-matrix derived primarily from 
United States data represents the cognitive structure of responses for students in a South African context. 
The Q-matrix for this study is shown in Table \ref{tab:LeQ} which shows the one developed by \citet{le:2025} and also shows 
(with arrows: $\longrightarrow$) where we found slightly different item attributes: $19,26$ and $27$. In what follows 
these differences make little change to the item DIF analyses and also in terms of Q-matrix validation checks that we performed.

\subsubsection{Multi-Skill Item Structure} 

\noindent As we discussed, analysis of the FCI suggests that a number of items require the simultaneous coordination of 
multiple skills (e.g. vector reasoning, 
conceptual relationships, and visualization), rather than a single dominant attribute, see Table \ref{table:skills}. 
This is reflected in slight deviations that we found from previously published Q-matrices (see Table \ref{tab:LeQ}) where both 
versions lend support to the notion that various items have a multi-skill structure (Table \ref{tab:LeQ}). This motivates the  
use of higher-order CDMs (like G-DINA) and reinforces the need to evaluate whether a given Q-matrix remains 
valid across contexts. 

\subsubsection{DIF in Cognitive Diagnostic Models} 

\noindent Within the CDM framework, DIF is naturally defined at the attribute level. For a discussion of such approaches within the DINA 
framework see \citet{hou2014differential} and for further generalizations see
\citet{svetina_examining_2018, li_assessment_2015, martinkova_checking_2017, Qiu:2019}. 
For two 
groups $g \in \{\mathrm{US}, \mathrm{SA}\}$, DIF occurs for item $j$ if 
\[ P(Y_{ij}=1 \mid \boldsymbol{\alpha}_i, g=\mathrm{US}) \;\neq\; P(Y_{ij}=1 \mid 
\boldsymbol{\alpha}_i, g=\mathrm{SA}), 
\] 
for at least one attribute profile $\boldsymbol{\alpha}_i$. Equivalently, DIF can be expressed in terms of item parameters, 
see Figure \ref{fig:DIFCDM}. For example, under a DINA-type parameterization, 
\begin{equation}
(s_j^{\mathrm{US}}, g_j^{\mathrm{US}}) \;\neq\; (s_j^{\mathrm{SA}}, g_j^{\mathrm{SA}}), 
\end{equation} 
or more  generally under G-DINA \cite[e.g., see][]{ma_gdina_2020}, 
\[ 
\boldsymbol{\delta}_j^{\mathrm{US}} \neq \boldsymbol{\delta}_j^{\mathrm{SA}}, 
\] 
where $\boldsymbol{\delta}_j$ denotes the vector of item response function parameters associated with latent attribute combinations. 
Some sample slip and guess parameters are shown in Table \ref{tab:supp_item_params} based on an analysis of US and South African data.

Statistical tests of DIF are therefore conducted by testing the null hypothesis 
\[ H_0: 
\boldsymbol{\delta}_j^{\mathrm{US}} = \boldsymbol{\delta}_j^{\mathrm{SA}}. \] 
Rejection of $H_0$ indicates that the probability structure of item $j$ differs between groups even after conditioning on attribute 
mastery, providing direct evidence of non-invariance in the cognitive model. To test $H_0$, we calculate the Wald statistic within the 
G-DINA framework \cite{hou2014differential, delaTorre:2009}. Multiple 
comparisons are controlled using the Benjamini-Hochberg False Discovery Rate (FDR) procedure ($p_{\mathrm{adj}} < 0.05$). Additionally, 
Cohen's $h$ effect sizes can be calculated for item difficulty differences between the UJ ($N = 1,016$) and LASSO ($N = 4,750$) cohorts, 
as we explain. 

\subsubsection{Effect Size Measures for Differential Item Functioning}
\noindent To quantify the magnitude of item-level differences between the United States (LASSO) and South African (UJ) cohorts, 
we considered Cohen's $h$ effect size rather than Cohen's $d$ \cite[e.g., see][for a discussion of Cohen's $d$ for gains in A 
PER context]{Nissen:2018} and \citep{kraft:2020-effect-size} for that in education in general.

For two proportions, \citet[][chap. 5]{cohen_statistical_2013} proposed the effect size 
\begin{equation} h = 
2\arcsin(\sqrt{P_1}) - 2\arcsin(\sqrt{P_2}), 
\end{equation} 
which applies the arcsine-square-root transformation to stabilize the variance of proportions. 
Unlike Cohen's $d$, Cohen's $h$ is specifically designed for comparisons between proportions and therefore 
provides a more robust measure of DIF when item performance is expressed as success rates. 
The conventional interpretation thresholds are: 
$$
|h| < 0.20 \qquad \text{(Negligible)},  
\qquad 0.20 \le |h| <  0.50 \qquad \text{(Moderate)}, 
\qquad |h| \ge 0.50 \qquad \text{(High)}
$$
where our checks have shown that Cohen's $d$ over inflates the effect size for items with proportional success rates, see 
Table \ref{tab:step2_dif_comparison}.

\subsubsection{Interpretation for Q-matrix Transferability} 

\noindent In this framework, DIF serves as a diagnostic tool for assessing Q-matrix transferability. 
If a ``transported'' Q-matrix is valid across contexts, then conditional item probabilities should be invariant across groups. 
Conversely, systematic 
DIF indicates that the Q-matrix fails to fully capture the cognitive processes underlying item responses in one or both 
contexts. We therefore interpret DIF not merely as a psychometric artifact but as evidence of potential mismatches in 
the assumed cognitive structure of the FCI across educational settings.

\section{Data Sources and Sample Preparation}
\label{sec:data_sources}

To evaluate item-level measurement invariance and Q-matrix transferability across distinct educational contexts, 
response data were compiled from two independent international cohorts: the United States LASSO database and the South African 
UJ dataset.

\subsection{LASSO Cohort (US)}
\label{sec:LASSO}

\noindent The baseline dataset was extracted from the LASSO portal \citep{lasso:nodate, nissen_providing_2022}, 
comprising $N_{\text{raw}} = 7,129$ student entries across multiple US higher education institutions (first three years). 
Responses were filtered for pre-test administration to make a similar comparison with the UJ cohort. 
Cases with incomplete item-level data required for G-DINA parameter estimation were omitted, 
yielding a final analyzed US sample of $N_{\text{LASSO}} = 4,750$.

\subsection{University of Johannesburg Cohort (South Africa)}
\label{sec:UJ}

\noindent The South African dataset consists of introductory physics student assessments collected at UJ between 
2020 and 2024, \citep[][e.g., see for a discussion of the cohorts]{Chrysostomou:2024}. 
Following data cleaning to isolate unique, valid pre-test submissions (or first attempts where pre-tests were absent), 
the final analyzed cohort comprises $N_{\text{UJ}} = 1,016$ students. It is worth mentioning that the demographics in 
South Africa are varied multi-culturally and linguistically with 11 official spoken languages \citep[e.g., see][]{Chrysostomou:2024}

\subsection{The FCI Q-matrix Across Continents}
\label{sec:qmat}

\noindent In the section we discuss how we initially defined the Q-matrix used in this work and 
mention how it compares to that of the 
work used to study LASSO data in the US using CDM models \citep{le:2025}. Agreement was broadly found with the attribute
definitions 
proposed by \citet{le:2025}, particularly with respect to the overarching skills associated with vector reasoning, conceptual 
relationships, and visualization, see Table \ref{tab:LeQ}. However, our examination of the FCI items suggested that item 
questions $19, 26$ and $27$ might be adequately characterized by more than one single attribute.\footnote{Note that in 
\citep{le:2025} and in our own work, one can run Q-matrix validation checks for a suggested Q-matrix that gives a better fit 
(see the $*$'s in Table \ref{tab:LeQ} as examples).}

As we shall see in the next section, 
using either Q-matrix, we found agreement between the group differences 
for the US LASSO and the South African UJ cohorts, apart from a few differences in 
effect-size, but with no major-shifts in DIF severity 
classification, see the Results section. 
This comparison provides a starting point for examining Q-matrix transferability 
across educational contexts and assessing whether differences in item functioning arise 
from alternative cognitive structures or 
from genuine differences in student reasoning. 

\section{Results and Analyses}
\label{sec:res}

\subsection*{Q-Matrix Coding for the FCI}
\label{sec:qcode}

\noindent We will now discuss the coding of the Q-matrix, where the skills are defined as: apply vectors ($k_1$); conceptual 
relationships ($k_2$); visual relationships ($k_3$), see Table \ref{table:skills}. Note that based on the framework of 
\cite{le:2025} which uses a suit of different mechanics cognitive diagnostics there are more skills used in that work 
\citep{le:2025}. Our approach differs in that we do not fully align with their Q-matrix, 
which we tested using different DINA models (see Table~\ref{tab:LeQ}).

To evaluate the transferability of the FCI cognitive structure across distinct educational contexts, independent 
Generalized DINA (G-DINA) models were calibrated on the US LASSO cohort ($N = 4,750$) and the South African University of 
Johannesburg (UJ) cohort ($N = 1,016$). Item parameters were estimated using maximum likelihood with 2,000 EM iterations. 
Model fit was evaluated using $M_2$ statistics and item-level RMSEA values ($\text{RMSEA} < 0.045$ across both cohorts), 
confirming adequate absolute model fit for both datasets under the generalized framework.

\subsection{Differential Item Functioning (DIF) Analysis}
\label{sec:DIF}

\begin{figure}[H]
\begin{minipage}{\linewidth}
    \centering
   \caption[]{
   FCI Differential Item Functioning (DIF) effect sizes (Cohen's $h$) comparing UJ ($N=1,016$) and LASSO ($N=4,750$) 
  cohorts\footnote{Dotted and dashed lines indicate moderate ($h \ge 0.2$) and high ($h \ge 0.5$) DIF thresholds, respectively. 
  Confidence intervals for each item are shown as black horizontal lines.}
  }
  \label{fig:fci_dif_plot}
 \includegraphics[width=0.8\linewidth]{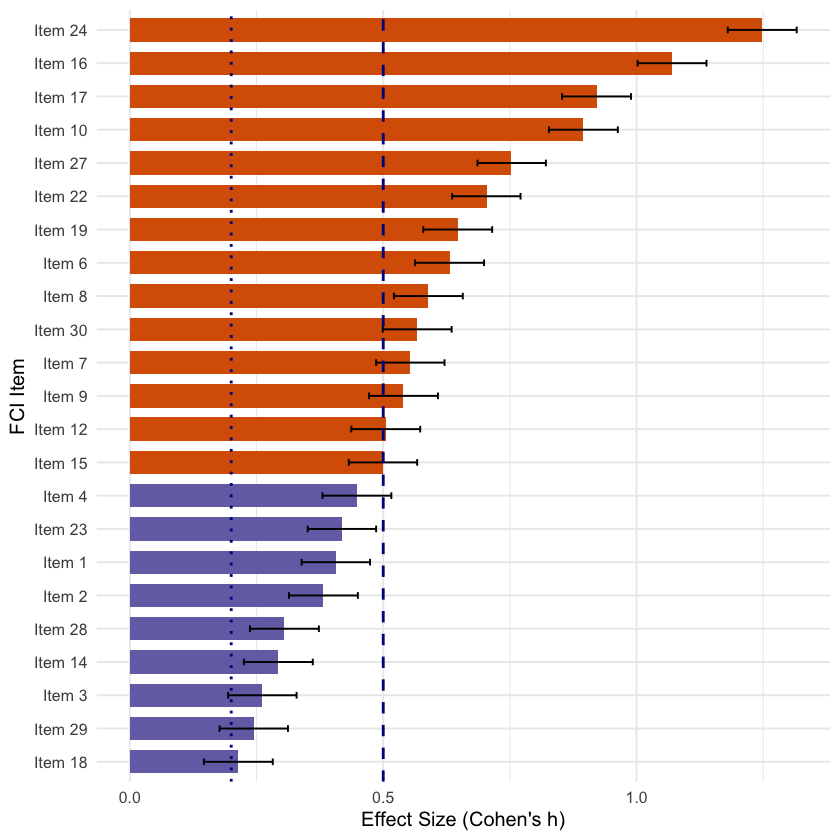}
 \end{minipage}
\end{figure}

\noindent To test item-level measurement invariance between the US and South African cohorts, item-level Wald 
tests were executed 
within the G-DINA framework \citep{hou2014differential} (see also \citet{svetina_examining_2018, li_assessment_2015, 
martinkova_checking_2017} for further discussions of DIF analyses in CDMs). 
$p$-values were adjusted for multiple testing using the Benjamini--Hochberg False Discovery Rate (FDR) procedure 
($p_{\text{adj}} < 0.05$). Additionally, Cohen's $h$ effect sizes were calculated for proportional success rates 
($P^+$) to categorize DIF severity into Negligible ($|h| < 0.2$), Moderate ($0.2 \le |h| < 0.5$), and High DIF ($|h| \ge 
0.5$). Figure~\ref{fig:fci_dif_plot} displays the distribution of effect sizes across all FCI items, highlighting the  
parameter shifts observed between cohorts. Table~\ref{tab:fci_dif_cohen_h} summarizes the item difficulties, Wald 
statistics, adjusted $p$-values, effect sizes, and DIF classifications across all 30 items.

\begin{table}[H]
\centering
\caption{FCI Differential Item Functioning Results using Cohen's $h$} 
\label{tab:fci_dif_cohen_h}
\resizebox{0.8\textwidth}{!}{
\begin{threeparttable}
\begin{tabular}{lrrrrrrl}
  \hline
Item & $^\dagger P_{\text{LASSO}}$ & $^\dagger P_{\text{UJ}}$ & adj.pvalue & $\rm Cohen_h$ & $h_{\text{lower}}$ & $h_{\text{upper}}$ & 
$\rm DIF_{severity}$ \\ 
  \hline
 24 & 0.63 & 0.08 & < 0.001 & 1.25 & 1.18 & 1.32 & High DIF (**) \\ 
 16 & 0.56 & 0.10 & < 0.001 & 1.07 & 1.00 & 1.14 & High DIF (**) \\ 
 17 & 0.14 & 0.56 & < 0.001 & 0.92 & 0.85 & 0.99 & High DIF (**) \\ 
 10 & 0.51 & 0.12 & < 0.001 & 0.90 & 0.83 & 0.96 & High DIF (**) \\ 
 27 & 0.57 & 0.21 & < 0.001 & 0.75 & 0.69 & 0.82 & High DIF (**) \\ 
 22 & 0.42 & 0.12 & < 0.001 & 0.70 & 0.64 & 0.77 & High DIF (**) \\ 
 19 & 0.06 & 0.28 & < 0.001 & 0.65 & 0.58 & 0.71 & High DIF (**) \\ 
 6  & 0.73 & 0.43 & < 0.001 & 0.63 & 0.56 & 0.70 & High DIF (**) \\ 
 8  & 0.60 & 0.31 & < 0.001 & 0.59 & 0.52 & 0.66 & High DIF (**) \\ 
 30 & 0.23 & 0.04 & < 0.001 & 0.57 & 0.50 & 0.64 & High DIF (**) \\ 
 7  & 0.62 & 0.34 & < 0.001 & 0.55 & 0.49 & 0.62 & High DIF (**) \\ 
 9  & 0.07 & 0.26 & < 0.001 & 0.54 & 0.47 & 0.61 & High DIF (**) \\ 
 12 & 0.67 & 0.42 & < 0.001 & 0.51 & 0.44 & 0.57 & High DIF (**) \\ 
 15 & 0.29 & 0.10 & < 0.001 & 0.50 & 0.43 & 0.57 & High DIF (**) \\ 
 4  & 0.32 & 0.54 & < 0.001 & 0.45 & 0.38 & 0.52 & Moderate DIF (*) \\ 
 23 & 0.20 & 0.06 & < 0.001 & 0.42 & 0.35 & 0.49 & Moderate DIF (*) \\ 
 1  & 0.67 & 0.47 & < 0.001 & 0.41 & 0.34 & 0.47 & Moderate DIF (*) \\ 
 2  & 0.35 & 0.19 & < 0.001 & 0.38 & 0.31 & 0.45 & Moderate DIF (*) \\ 
 25 & 0.20 & 0.09 & 0.170 & 0.32 & 0.25 & 0.39 & Negligible DIF \\ 
 28 & 0.33 & 0.47 & < 0.001 & 0.30 & 0.24 & 0.37 & Moderate DIF (*) \\ 
 14 & 0.42 & 0.28 & < 0.001 & 0.29 & 0.23 & 0.36 & Moderate DIF (*) \\ 
 3  & 0.49 & 0.37 & < 0.001 & 0.26 & 0.19 & 0.33 & Moderate DIF (*) \\ 
 29 & 0.45 & 0.57 & < 0.001 & 0.24 & 0.18 & 0.31 & Moderate DIF (*) \\ 
 20 & 0.39 & 0.28 & 0.170 & 0.23 & 0.16 & 0.30 & Negligible DIF \\ 
 18 & 0.20 & 0.29 & < 0.001 & 0.21 & 0.15 & 0.28 & Moderate DIF (*) \\ 
 26 & 0.10 & 0.17 & < 0.001 & 0.19 & 0.12 & 0.26 & Negligible DIF \\ 
 21 & 0.18 & 0.14 & < 0.001 & 0.13 & 0.06 & 0.20 & Negligible DIF \\ 
 13 & 0.22 & 0.26 & < 0.001 & 0.09 & 0.02 & 0.15 & Negligible DIF \\ 
 5  & 0.17 & 0.14 & < 0.001 & 0.09 & 0.02 & 0.15 & Negligible DIF \\ 
 11 & 0.20 & 0.17 & < 0.001 & 0.08 & 0.01 & 0.15 & Negligible DIF \\ 
\hline
\end{tabular}
\vskip 0.2cm
\begin{tablenotes}
      \footnotesize
      \item[$\dagger$] $P_{\text{LASSO}}$ and $P_{\text{UJ}}$ denote the success rates for each cohort, respectively.
      \item[*] Table 2 reports absolute effect size magnitudes ($|h|$) rounded to two decimal places alongside Benjamini--Hochberg adjusted $p$-values ($p_{\text{adj}}$). Directional effect sizes showing cohort performance advantages and exact three-decimal values are provided in Appendix Table D1.
    \end{tablenotes}
\end{threeparttable}
}
\end{table}

\subsection{Cross-Cultural Cognitive Differences}
\label{sec:crosscult}

\noindent Out of 30 items, 14 exhibited High DIF ($|h| \ge 0.50$), 9 exhibited Moderate DIF ($0.20 \le |h| < 0.50$), and 7 exhibited Negligible DIF ($|h| < 0.20$). We do not intend to unpack the physics behind each item where we refer the reader to \citep[e.g., see][]{Hest:1992,chrysostomou:2025, Bani_Salameh:2016a}. The analysis suggests distinct conceptual patterns between the US and South African student populations:

\begin{enumerate}
    \item \textbf{Continuous Force Dynamics (LASSO Advantage):} Items involving continuous forces, kinetic friction, and dynamic vector acceleration showed extreme DIF favoring the US cohort. Specifically, \textbf{Item 24} ($62.7\%$ vs. $8.2\%$, $h = 1.25$) and \textbf{Item 16} ($56.5\%$ vs. $9.6\%$, $h = 1.07$) showed the largest performance gaps, suggesting differences in instructional coverage or problem representation in dynamic mechanics.
    \item \textbf{Elevator Contact Forces (UJ Advantage):} Conversely, \textbf{Item 17} (evaluating forces in an elevator system) demonstrated a strong performance advantage for the UJ cohort ($55.9\%$ vs. $14.0\%$, $h = 0.92$). This suggests specific context-dependent reasoning or localized curricular emphasis on normal force equilibrium during acceleration.
\end{enumerate}

\subsection{Q-Matrix Sensitivity Analysis}
\noindent To verify whether the observed DIF flags stem from true latent trait differences rather than Q-matrix specification artifacts, the entire estimation pipeline was re-evaluated using the specification established by \cite{le:2025}. The DIF severity classifications proved $100\%$ invariant across both Q-matrix structures: zero items shifted between severity categories ($14$ High DIF, $9$ Moderate DIF, and $7$ Negligible DIF under both specifications). While minor fluctuations in Wald statistics occurred due to attribute re-parameterization, every item's adjusted $p$-value remained significant ($p_{\text{adj}} < 0.001$). Key High-DIF indicators: most notably Item 24 ($h = 1.25$), Item 16 ($h = 1.07$), and Item 17 ($h = 0.92$); exhibited identical effect size magnitudes across both Q-matrices. 

It is worth mentioning that this structural measurement invariance may well reflect cross-cultural cognitive differences in physics reasoning rather than model specification bias.
\section{Conclusion}
\label{sec:conc}

\noindent This study demonstrates that Q-matrix transferability across international higher education contexts cannot be 
assumed when applying CDMs to research-based assessments like the FCI. Although independent G-DINA 
calibrations established adequate absolute fit ($\text{RMSEA} < 0.045$) for both the US (LASSO) and South African (UJ) 
cohorts individually, cross-cultural DIF analysis uncovered cognitive differences; particularly in continuous force 
dynamics and contact force reasoning. 

To answer our initial research questions:

\begin{enumerate}
 \item Can an FCI Q-matrix developed from United States cohorts be validly transported to a South African higher 
 education context?
 \begin{itemize}
     \item \it Yes, the FCI Q-matrix developed by \citet{le:2025} gives the same DIF analysis results as 
     compared to a slightly different Q-matrix we used, see Table \ref{tab:LeQ}; within the assumption of using the G-DINA 
     model, see Table \ref{tab:fci_dif_cohen_h}.
 \end{itemize}
 \item When comparing DIF between US and South African cohorts are there any 
 significant effect-sizes for items in the FCI?
    \begin{itemize}
        \item \it Yes, different items had different effect-sizes for 
        different questions suggesting that beyond Q-matrix comparability these differences stem from cultural-contexts in 
        the approach to solving those particular items (see Figure \ref{fig:fci_dif_plot} and Table \ref{tab:fci_dif_cohen_h}).
    \end{itemize}
\end{enumerate}

Crucially, our sensitivity analyses demonstrated that these diagnostic differences are entirely invariant to the underlying 
Q-matrix specification. Because all $30$ items retained identical DIF severity classifications when evaluated under both the 
UJ-modified Q-matrix and the one used in \citet{le:2025}. This leads to the suggestion that these cognitive differences 
represent authentic cross-cultural differences in student reasoning rather than psychometric artifacts; this issue requires 
further investigation.

Several limitations should be noted. Like all CDM approaches, the G-DINA framework is dependent on the 
specification of the Q-matrix and the theoretical assumptions used to define the underlying attributes. Although empirical 
Q-matrix validation and alternative specifications were explored, other plausible attribute structures may exist and could yield 
different interpretations of student mastery profiles. Furthermore, while the large cohort sizes analyzed in this study 
($N = 4,750$ and $N = 1,016$) 
support stable parameter estimation, CDMs remain sensitive to sparse attribute patterns and unobserved contextual 
factors. The present analysis is also restricted to students from the United States and South Africa, limiting the 
generalization of the findings to other educational settings. Nevertheless, the consistency of the results across alternative Q-
matrix specifications, together 
with adequate G-DINA model fits (unpublished) and invariant DIF classifications under sensitivity 
analyses (see Table \ref{tab:fci_dif_cohen_h}) suggests 
that the reported cross-cultural differences reflect substantive differences in student reasoning 
rather than artifacts of a particular model specification.

More broadly, these findings highlight the importance of evaluating measurement invariance before transferring 
research-based assessments across educational, cultural, and institutional contexts. While the FCI remains a robust 
instrument for probing conceptual understanding, the results suggest that the cognitive pathways students use to 
arrive at responses may differ across cohorts. Consequently, assumptions regarding universal cognitive structures should be tested 
rather than presumed. 

The G-DINA framework provides a powerful approach to such investigations by combining fine-grained 
diagnostic information with explicit testing of cross-group comparability. Future work will extend this approach to 
other concept inventories such as the Brief Electricity and Magnetism Assessment (BEMA), e.g., see \cite{Hansen-Stewart:2021}, 
investigate attribute-level mastery profiles 
in greater detail, and explore the development of adaptive diagnostic assessments capable of providing more fine-grained 
feedback to diverse student populations and to look at cohorts from other localities.
This study hopefully challenges assumptions about cognitive invariance in PER instruments and has 
implications for future adaptive and equity-focused assessment design.

\section{Acknowledgments}

We would like to thank Ben Van Dusen, Vy Le (Iowa State University), and Anna Chrysostomou (Sorbonne University) 
for valuable discussions, as well as the LASSO team for the use of the three year dataset used in 
this work. ASC was partly supported by the National Research Foundation of South Africa.


\bibliography{references.bib} 

\begin{thebibliography}{}

\bibitem [\protect \citeauthoryear {%
Adams%
, Hean%
, Sturgis%
\BCBL {}\ \BBA {} Clark%
}{%
Adams%
\ \protect \BOthers {.}}{%
{\protect \APACyear {2006}}%
}]{%
Adams:2006}
\APACinsertmetastar {%
Adams:2006}%
\begin{APACrefauthors}%
Adams, K.%
, Hean, S.%
, Sturgis, P.%
\BCBL {}\ \BBA {} Clark, J\BPBI M.%
\end{APACrefauthors}%
\unskip\
\newblock
\APACrefYearMonthDay{2006}{}{}.
\newblock
{\BBOQ}\APACrefatitle {Investigating the factors influencing professional
  identity of first-year health and social care students} {Investigating the
  factors influencing professional identity of first-year health and social
  care students}.{\BBCQ}
\newblock
\APACjournalVolNumPages{Learning in Health and Social Care}{5}{2}{55-68}.
\newblock
\begin{APACrefURL}
  \url{https://onlinelibrary.wiley.com/doi/abs/10.1111/j.1473-6861.2006.00119.x}
  \end{APACrefURL}
\newblock
\begin{APACrefDOI} \doi{https://doi.org/10.1111/j.1473-6861.2006.00119.x}
  \end{APACrefDOI}
\PrintBackRefs{\CurrentBib}

\bibitem [\protect \citeauthoryear {%
Affriyenni%
, Georgiou%
\BCBL {}\ \BBA {} Finkelstein%
}{%
Affriyenni%
\ \protect \BOthers {.}}{%
{\protect \APACyear {2025}}%
}]{%
affriyenni:2025}
\APACinsertmetastar {%
affriyenni:2025}%
\begin{APACrefauthors}%
Affriyenni, Y.%
, Georgiou, H.%
\BCBL {}\ \BBA {} Finkelstein, N.%
\end{APACrefauthors}%
\unskip\
\newblock
\APACrefYearMonthDay{2025}{{\APACmonth{09}}}{}.
\newblock
{\BBOQ}\APACrefatitle {Navigating the adoption of research-based instructional
  strategies within the complex nature of higher education} {Navigating the
  adoption of research-based instructional strategies within the complex nature
  of higher education}.{\BBCQ}
\newblock
\APACjournalVolNumPages{Phys. Rev. Phys. Educ. Res.}{21}{2}{020124}.
\newblock
\begin{APACrefURL}
  [{2026-09-10}]\url{https://link.aps.org/doi/10.1103/wvjk-8y8k}
  \end{APACrefURL}
\newblock
\begin{APACrefDOI} \doi{10.1103/wvjk-8y8k} \end{APACrefDOI}
\PrintBackRefs{\CurrentBib}

\bibitem [\protect \citeauthoryear {%
Allen%
\ \BBA {} Ross%
}{%
Allen%
\ \BBA {} Ross%
}{%
{\protect \APACyear {2017}}%
}]{%
allen_ross:2017}
\APACinsertmetastar {%
allen_ross:2017}%
\begin{APACrefauthors}%
Allen, G\BPBI D.%
\BCBT {}\ \BBA {} Ross, A.%
\end{APACrefauthors}%
\ (\BEDS).
\unskip\
\newblock
\APACrefYear{2017}.
\newblock
\APACrefbtitle {Pedagogy and {Content} in {Middle} and {High} {School}
  {Mathematics}} {Pedagogy and {Content} in {Middle} and {High} {School}
  {Mathematics}}.
\newblock
\APACaddressPublisher{Rotterdam}{SensePublishers}.
\newblock
\begin{APACrefURL}
  [{2024-10-04}]\url{http://link.springer.com/10.1007/978-94-6351-137-7}
  \end{APACrefURL}
\newblock
\begin{APACrefDOI} \doi{10.1007/978-94-6351-137-7} \end{APACrefDOI}
\PrintBackRefs{\CurrentBib}

\bibitem [\protect \citeauthoryear {%
Bani-Salameh%
}{%
Bani-Salameh%
}{%
{\protect \APACyear {2016}}%
}]{%
Bani_Salameh:2016a}
\APACinsertmetastar {%
Bani_Salameh:2016a}%
\begin{APACrefauthors}%
Bani-Salameh, H\BPBI N.%
\end{APACrefauthors}%
\unskip\
\newblock
\APACrefYearMonthDay{2016}{dec}{}.
\newblock
{\BBOQ}\APACrefatitle {How persistent are the misconceptions about force and
  motion held by college students?} {How persistent are the misconceptions
  about force and motion held by college students?}{\BBCQ}
\newblock
\APACjournalVolNumPages{Physics Education}{52}{1}{014003}.
\newblock
\begin{APACrefURL} \url{https://doi.org/10.1088/1361-6552/52/1/014003}
  \end{APACrefURL}
\newblock
\begin{APACrefDOI} \doi{10.1088/1361-6552/52/1/014003} \end{APACrefDOI}
\PrintBackRefs{\CurrentBib}

\bibitem [\protect \citeauthoryear {%
Birnbaum%
}{%
Birnbaum%
}{%
{\protect \APACyear {1968}}%
}]{%
Birnbaum:1968}
\APACinsertmetastar {%
Birnbaum:1968}%
\begin{APACrefauthors}%
Birnbaum, A.%
\end{APACrefauthors}%
\unskip\
\newblock
\APACrefYearMonthDay{1968}{}{}.
\newblock
{\BBOQ}\APACrefatitle {Some latent trait models and their use in inferring an
  examinee’s ability} {Some latent trait models and their use in inferring an
  examinee’s ability}.{\BBCQ}
\newblock
\BIn{} F\BPBI M.~Lord\ \BBA {} M\BPBI R.~Novick\ (\BEDS), \APACrefbtitle
  {Statistical Theories of Mental Test Scores} {Statistical theories of mental
  test scores}\ (\BPGS\ 397--472).
\newblock
\APACaddressPublisher{}{Addison-Wesley}.
\PrintBackRefs{\CurrentBib}

\bibitem [\protect \citeauthoryear {%
Buncher%
, Nissen%
, Van~Dusen%
\BCBL {}\ \BBA {} Talbot%
}{%
Buncher%
\ \protect \BOthers {.}}{%
{\protect \APACyear {2025}}%
}]{%
buncher:2025}
\APACinsertmetastar {%
buncher:2025}%
\begin{APACrefauthors}%
Buncher, J\BPBI B.%
, Nissen, J\BPBI M.%
, Van~Dusen, B.%
\BCBL {}\ \BBA {} Talbot, R\BPBI M.%
\end{APACrefauthors}%
\unskip\
\newblock
\APACrefYearMonthDay{2025}{{\APACmonth{04}}}{}.
\newblock
{\BBOQ}\APACrefatitle {Is the {Force} {Concept} {Inventory} biased across the
  intersections of gender and race?} {Is the {Force} {Concept} {Inventory}
  biased across the intersections of gender and race?}{\BBCQ}
\newblock
\APACjournalVolNumPages{Phys. Rev. Phys. Educ. Res.}{21}{1}{010137}.
\newblock
\begin{APACrefURL}
  [{2026-03-31}]\url{https://link.aps.org/doi/10.1103/PhysRevPhysEducRes.21.010137}
  \end{APACrefURL}
\newblock
\begin{APACrefDOI} \doi{10.1103/PhysRevPhysEducRes.21.010137} \end{APACrefDOI}
\PrintBackRefs{\CurrentBib}

\bibitem [\protect \citeauthoryear {%
Chrysostomou%
, Cornell%
\BCBL {}\ \BBA {} Naylor%
}{%
Chrysostomou%
\ \protect \BOthers {.}}{%
{\protect \APACyear {2024}}%
}]{%
Chrysostomou:2024}
\APACinsertmetastar {%
Chrysostomou:2024}%
\begin{APACrefauthors}%
Chrysostomou, A.%
, Cornell, A\BPBI S.%
\BCBL {}\ \BBA {} Naylor, W.%
\end{APACrefauthors}%
\unskip\
\newblock
\APACrefYearMonthDay{2024}{}{}.
\newblock
{\BBOQ}\APACrefatitle {A three-year comparative study of dominant
  misconceptions among first-year physics students at a South African
  university} {A three-year comparative study of dominant misconceptions among
  first-year physics students at a south african university}.{\BBCQ}
\newblock
\APACjournalVolNumPages{Physics Education}{59}{1}{015036}.
\newblock
\begin{APACrefURL} \url{https://dx.doi.org/10.1088/1361-6552/ad14ea}
  \end{APACrefURL}
\newblock
\begin{APACrefDOI} \doi{10.1088/1361-6552/ad14ea} \end{APACrefDOI}
\PrintBackRefs{\CurrentBib}

\bibitem [\protect \citeauthoryear {%
Chrysostomou%
, Cornell%
\BCBL {}\ \BBA {} Naylor%
}{%
Chrysostomou%
\ \protect \BOthers {.}}{%
{\protect \APACyear {2025}}%
}]{%
chrysostomou:2025}
\APACinsertmetastar {%
chrysostomou:2025}%
\begin{APACrefauthors}%
Chrysostomou, A.%
, Cornell, A\BPBI S.%
\BCBL {}\ \BBA {} Naylor, W.%
\end{APACrefauthors}%
\unskip\
\newblock
\APACrefYearMonthDay{2025}{}{}.
\newblock
{\BBOQ}\APACrefatitle {Dominant misconceptions and alluvial flows between
  Engineering and Physical Science students} {Dominant misconceptions and
  alluvial flows between engineering and physical science students}.{\BBCQ}
\newblock
\APACjournalVolNumPages{Physics Education}{60}{3}{035026}.
\newblock
\begin{APACrefURL} \url{https://doi.org/10.1088/1361-6552/adc4ad}
  \end{APACrefURL}
\newblock
\begin{APACrefDOI} \doi{10.1088/1361-6552/adc4ad} \end{APACrefDOI}
\PrintBackRefs{\CurrentBib}

\bibitem [\protect \citeauthoryear {%
Cohen%
}{%
Cohen%
}{%
{\protect \APACyear {1988}}%
}]{%
cohen_statistical_2013}
\APACinsertmetastar {%
cohen_statistical_2013}%
\begin{APACrefauthors}%
Cohen, J.%
\end{APACrefauthors}%
\unskip\
\newblock
\APACrefYear{1988}.
\newblock
\APACrefbtitle {Statistical {Power} {Analysis} for the {Behavioral} {Sciences}}
  {Statistical {Power} {Analysis} for the {Behavioral} {Sciences}}\
  (\PrintOrdinal{2nd}\ \BEd).
\newblock
\APACaddressPublisher{}{Routledge}.
\newblock
\begin{APACrefURL} \url{https://www.taylorfrancis.com/books/9781134742707}
  \end{APACrefURL}
\newblock
\begin{APACrefDOI} \doi{10.4324/9780203771587} \end{APACrefDOI}
\PrintBackRefs{\CurrentBib}

\bibitem [\protect \citeauthoryear {%
de~la Torre%
}{%
de~la Torre%
}{%
{\protect \APACyear {2009}}%
}]{%
delaTorre:2009}
\APACinsertmetastar {%
delaTorre:2009}%
\begin{APACrefauthors}%
de~la Torre, J.%
\end{APACrefauthors}%
\unskip\
\newblock
\APACrefYearMonthDay{2009}{{\APACmonth{05}}}{}.
\newblock
{\BBOQ}\APACrefatitle {A {Cognitive} {Diagnosis} {Model} for {Cognitively}
  {Based} {Multiple}-{Choice} {Options}} {A {Cognitive} {Diagnosis} {Model} for
  {Cognitively} {Based} {Multiple}-{Choice} {Options}}.{\BBCQ}
\newblock
\APACjournalVolNumPages{Applied Psychological Measurement}{33}{3}{163--183}.
\newblock
\begin{APACrefURL}
  [{2024-09-25}]\url{https://journals.sagepub.com/doi/10.1177/0146621608320523}
  \end{APACrefURL}
\newblock
\begin{APACrefDOI} \doi{10.1177/0146621608320523} \end{APACrefDOI}
\PrintBackRefs{\CurrentBib}

\bibitem [\protect \citeauthoryear {%
de~la Torre%
}{%
de~la Torre%
}{%
{\protect \APACyear {2011}}%
}]{%
delaTorre:2011}
\APACinsertmetastar {%
delaTorre:2011}%
\begin{APACrefauthors}%
de~la Torre, J.%
\end{APACrefauthors}%
\unskip\
\newblock
\APACrefYearMonthDay{2011}{}{}.
\newblock
{\BBOQ}\APACrefatitle {The generalized {DINA} model framework} {The generalized
  {DINA} model framework}.{\BBCQ}
\newblock
\APACjournalVolNumPages{Psychometrika}{76}{}{179--199}.
\newblock
\begin{APACrefDOI} \doi{https://doi.org/10.1007/s11336-011-9207-7}
  \end{APACrefDOI}
\PrintBackRefs{\CurrentBib}

\bibitem [\protect \citeauthoryear {%
Hake%
}{%
Hake%
}{%
{\protect \APACyear {1998}}%
}]{%
Hake:1998}
\APACinsertmetastar {%
Hake:1998}%
\begin{APACrefauthors}%
Hake, R\BPBI R.%
\end{APACrefauthors}%
\unskip\
\newblock
\APACrefYearMonthDay{1998}{}{}.
\newblock
{\BBOQ}\APACrefatitle {Interactive-engagement versus traditional methods: A
  six-thousand-student survey of mechanics test data for introductory physics
  courses} {Interactive-engagement versus traditional methods: A
  six-thousand-student survey of mechanics test data for introductory physics
  courses}.{\BBCQ}
\newblock
\APACjournalVolNumPages{American Journal of Physics}{66}{1}{64-74}.
\newblock
\begin{APACrefURL} \url{https://doi.org/10.1119/1.18809} \end{APACrefURL}
\newblock
\begin{APACrefDOI} \doi{10.1119/1.18809} \end{APACrefDOI}
\PrintBackRefs{\CurrentBib}

\bibitem [\protect \citeauthoryear {%
Halloun%
\ \BBA {} Hestenes%
}{%
Halloun%
\ \BBA {} Hestenes%
}{%
{\protect \APACyear {1985}}%
}]{%
Hall:1985}
\APACinsertmetastar {%
Hall:1985}%
\begin{APACrefauthors}%
Halloun, I\BPBI A.%
\BCBT {}\ \BBA {} Hestenes, D.%
\end{APACrefauthors}%
\unskip\
\newblock
\APACrefYearMonthDay{1985}{}{}.
\newblock
{\BBOQ}\APACrefatitle {The initial knowledge state of college physics students}
  {The initial knowledge state of college physics students}.{\BBCQ}
\newblock
\APACjournalVolNumPages{American Journal of Physics}{53}{}{1043}.
\PrintBackRefs{\CurrentBib}

\bibitem [\protect \citeauthoryear {%
Hansen%
\ \BBA {} Stewart%
}{%
Hansen%
\ \BBA {} Stewart%
}{%
{\protect \APACyear {2021}}%
}]{%
Hansen-Stewart:2021}
\APACinsertmetastar {%
Hansen-Stewart:2021}%
\begin{APACrefauthors}%
Hansen, J.%
\BCBT {}\ \BBA {} Stewart, J.%
\end{APACrefauthors}%
\unskip\
\newblock
\APACrefYearMonthDay{2021}{Nov}{}.
\newblock
{\BBOQ}\APACrefatitle {Multidimensional item response theory and the Brief
  Electricity and Magnetism Assessment} {Multidimensional item response theory
  and the brief electricity and magnetism assessment}.{\BBCQ}
\newblock
\APACjournalVolNumPages{Phys. Rev. Phys. Educ. Res.}{17}{}{020139}.
\newblock
\begin{APACrefURL}
  \url{https://link.aps.org/doi/10.1103/PhysRevPhysEducRes.17.020139}
  \end{APACrefURL}
\newblock
\begin{APACrefDOI} \doi{10.1103/PhysRevPhysEducRes.17.020139} \end{APACrefDOI}
\PrintBackRefs{\CurrentBib}

\bibitem [\protect \citeauthoryear {%
Henderson%
, Beach%
\BCBL {}\ \BBA {} Finkelstein%
}{%
Henderson%
\ \protect \BOthers {.}}{%
{\protect \APACyear {2011}}%
}]{%
Henderson2011}
\APACinsertmetastar {%
Henderson2011}%
\begin{APACrefauthors}%
Henderson, C.%
, Beach, A.%
\BCBL {}\ \BBA {} Finkelstein, N.%
\end{APACrefauthors}%
\unskip\
\newblock
\APACrefYearMonthDay{2011}{}{}.
\newblock
{\BBOQ}\APACrefatitle {Facilitating Change in Undergraduate STEM Instructional
  Practices: An Analytic Review of the Literature} {Facilitating change in
  undergraduate stem instructional practices: An analytic review of the
  literature}.{\BBCQ}
\newblock
\APACjournalVolNumPages{Journal of Research in Science
  Teaching}{48}{8}{952--984}.
\PrintBackRefs{\CurrentBib}

\bibitem [\protect \citeauthoryear {%
Hestenes%
}{%
Hestenes%
}{%
{\protect \APACyear {1998}}%
}]{%
Hest:1998}
\APACinsertmetastar {%
Hest:1998}%
\begin{APACrefauthors}%
Hestenes, D.%
\end{APACrefauthors}%
\unskip\
\newblock
\APACrefYearMonthDay{1998}{}{}.
\newblock
{\BBOQ}\APACrefatitle {Who needs physics education research!?} {Who needs
  physics education research!?}{\BBCQ}
\newblock
\APACjournalVolNumPages{American Journal of Physics}{66}{}{465}.
\PrintBackRefs{\CurrentBib}

\bibitem [\protect \citeauthoryear {%
Hestenes%
, Wells%
\BCBL {}\ \BBA {} Swackhamer%
}{%
Hestenes%
\ \protect \BOthers {.}}{%
{\protect \APACyear {1992}}%
}]{%
Hest:1992}
\APACinsertmetastar {%
Hest:1992}%
\begin{APACrefauthors}%
Hestenes, D.%
, Wells, M.%
\BCBL {}\ \BBA {} Swackhamer, G.%
\end{APACrefauthors}%
\unskip\
\newblock
\APACrefYearMonthDay{1992}{}{}.
\newblock
{\BBOQ}\APACrefatitle {Force concept inventory} {Force concept
  inventory}.{\BBCQ}
\newblock
\APACjournalVolNumPages{The Physics Teacher}{30}{3}{141-158}.
\newblock
\begin{APACrefURL} \url{https://doi.org/10.1119/1.2343497} \end{APACrefURL}
\newblock
\begin{APACrefDOI} \doi{10.1119/1.2343497} \end{APACrefDOI}
\PrintBackRefs{\CurrentBib}

\bibitem [\protect \citeauthoryear {%
Hou%
, de~la Torre%
\BCBL {}\ \BBA {} Nandakumar%
}{%
Hou%
\ \protect \BOthers {.}}{%
{\protect \APACyear {2014}}%
}]{%
hou2014differential}
\APACinsertmetastar {%
hou2014differential}%
\begin{APACrefauthors}%
Hou, L.%
, de~la Torre, J.%
\BCBL {}\ \BBA {} Nandakumar, R.%
\end{APACrefauthors}%
\unskip\
\newblock
\APACrefYearMonthDay{2014}{}{}.
\newblock
{\BBOQ}\APACrefatitle {Differential item functioning measurement in cognitive
  diagnosis models} {Differential item functioning measurement in cognitive
  diagnosis models}.{\BBCQ}
\newblock
\APACjournalVolNumPages{Journal of Educational and Behavioral
  Statistics}{39}{1}{21--44}.
\PrintBackRefs{\CurrentBib}

\bibitem [\protect \citeauthoryear {%
Kraft%
}{%
Kraft%
}{%
{\protect \APACyear {2020}}%
}]{%
kraft:2020-effect-size}
\APACinsertmetastar {%
kraft:2020-effect-size}%
\begin{APACrefauthors}%
Kraft, M\BPBI A.%
\end{APACrefauthors}%
\unskip\
\newblock
\APACrefYearMonthDay{2020}{}{}.
\newblock
{\BBOQ}\APACrefatitle {Interpreting Effect Sizes of Education Interventions}
  {Interpreting effect sizes of education interventions}.{\BBCQ}
\newblock
\APACjournalVolNumPages{Educational Researcher}{49}{4}{241-253}.
\newblock
\begin{APACrefDOI} \doi{10.3102/0013189X20912798} \end{APACrefDOI}
\PrintBackRefs{\CurrentBib}

\bibitem [\protect \citeauthoryear {%
{LASSO}%
}{%
{LASSO}%
}{%
{\protect \APACyear {{\protect \bibnodate {}}}}%
}]{%
lasso:nodate}
\APACinsertmetastar {%
lasso:nodate}%
\begin{APACrefauthors}%
{LASSO}.%
\end{APACrefauthors}%
\unskip\
\newblock
\APACrefYearMonthDay{{\protect \bibnodate {}}}{}{}.
\newblock
\APACrefbtitle {{LASSO} – {Learning} {About} {STEM} {Student} {Outcomes}.}
  {{LASSO} – {Learning} {About} {STEM} {Student} {Outcomes}.}
\newblock
\begin{APACrefURL} \url{https://lassoeducation.org/} \end{APACrefURL}
\PrintBackRefs{\CurrentBib}

\bibitem [\protect \citeauthoryear {%
Le%
, Nissen%
, Morphew%
, Chang%
\BCBL {}\ \BBA {} Dusen%
}{%
Le%
\ \protect \BOthers {.}}{%
{\protect \APACyear {2026}}%
}]{%
le:2026}
\APACinsertmetastar {%
le:2026}%
\begin{APACrefauthors}%
Le, V.%
, Nissen, J\BPBI M.%
, Morphew, J\BPBI W.%
, Chang, H\BPBI H.%
\BCBL {}\ \BBA {} Dusen, B\BPBI V.%
\end{APACrefauthors}%
\unskip\
\newblock
\APACrefYearMonthDay{2026}{}{}.
\newblock
{\BBOQ}\APACrefatitle {Mechanics Cognitive Diagnostic: Testing Fine-Grained
  Learning Objectives in Introductory Physics} {Mechanics cognitive diagnostic:
  Testing fine-grained learning objectives in introductory physics}.{\BBCQ}
\newblock
\APACjournalVolNumPages{arXiv}{}{}{}.
\newblock
\begin{APACrefURL} \url{https://arxiv.org/abs/2609.09584} \end{APACrefURL}
\PrintBackRefs{\CurrentBib}

\bibitem [\protect \citeauthoryear {%
Le%
\ \protect \BOthers {.}}{%
Le%
\ \protect \BOthers {.}}{%
{\protect \APACyear {2025}}%
}]{%
le:2025}
\APACinsertmetastar {%
le:2025}%
\begin{APACrefauthors}%
Le, V.%
, Nissen, J\BPBI M.%
, Tang, X.%
, Zhang, Y.%
, Mehrabi, A.%
, Morphew, J\BPBI W.%
\BDBL {}Van~Dusen, B.%
\end{APACrefauthors}%
\unskip\
\newblock
\APACrefYearMonthDay{2025}{}{}.
\newblock
{\BBOQ}\APACrefatitle {Applying cognitive diagnostic models to mechanics
  concept inventories} {Applying cognitive diagnostic models to mechanics
  concept inventories}.{\BBCQ}
\newblock
\APACjournalVolNumPages{Physical Review Physics Education
  Research}{21}{1}{010103}.
\newblock
\begin{APACrefURL}
  [{2025-09-28}]\url{https://link.aps.org/doi/10.1103/PhysRevPhysEducRes.21.010103}
  \end{APACrefURL}
\newblock
\begin{APACrefDOI} \doi{10.1103/PhysRevPhysEducRes.21.010103} \end{APACrefDOI}
\PrintBackRefs{\CurrentBib}

\bibitem [\protect \citeauthoryear {%
Li%
\ \BBA {} Wang%
}{%
Li%
\ \BBA {} Wang%
}{%
{\protect \APACyear {2015}}%
}]{%
li_assessment_2015}
\APACinsertmetastar {%
li_assessment_2015}%
\begin{APACrefauthors}%
Li, X.%
\BCBT {}\ \BBA {} Wang, W.%
\end{APACrefauthors}%
\unskip\
\newblock
\APACrefYearMonthDay{2015}{{\APACmonth{03}}}{}.
\newblock
{\BBOQ}\APACrefatitle {Assessment of {Differential} {Item} {Functioning}
  {Under} {Cognitive} {Diagnosis} {Models}: {The} {DINA} {Model} {Example}}
  {Assessment of {Differential} {Item} {Functioning} {Under} {Cognitive}
  {Diagnosis} {Models}: {The} {DINA} {Model} {Example}}.{\BBCQ}
\newblock
\APACjournalVolNumPages{J Educational Measurement}{52}{1}{28--54}.
\newblock
\begin{APACrefURL} \url{https://onlinelibrary.wiley.com/doi/10.1111/jedm.12061}
  \end{APACrefURL}
\newblock
\begin{APACrefDOI} \doi{10.1111/jedm.12061} \end{APACrefDOI}
\PrintBackRefs{\CurrentBib}

\bibitem [\protect \citeauthoryear {%
Ma%
\ \BBA {} De~La~Torre%
}{%
Ma%
\ \BBA {} De~La~Torre%
}{%
{\protect \APACyear {2020}}%
}]{%
ma_gdina_2020}
\APACinsertmetastar {%
ma_gdina_2020}%
\begin{APACrefauthors}%
Ma, W.%
\BCBT {}\ \BBA {} De~La~Torre, J.%
\end{APACrefauthors}%
\unskip\
\newblock
\APACrefYearMonthDay{2020}{}{}.
\newblock
{\BBOQ}\APACrefatitle {\textbf{{GDINA}} : {An} \textit{{R}} {Package} for
  {Cognitive} {Diagnosis} {Modeling}} {\textbf{{GDINA}} : {An} \textit{{R}}
  {Package} for {Cognitive} {Diagnosis} {Modeling}}.{\BBCQ}
\newblock
\APACjournalVolNumPages{J. Stat. Soft.}{93}{14}{}.
\newblock
\begin{APACrefURL} [{2026-09-03}]\url{http://www.jstatsoft.org/v93/i14/}
  \end{APACrefURL}
\newblock
\begin{APACrefDOI} \doi{10.18637/jss.v093.i14} \end{APACrefDOI}
\PrintBackRefs{\CurrentBib}

\bibitem [\protect \citeauthoryear {%
Madsen%
, McKagan%
\BCBL {}\ \BBA {} Sayre%
}{%
Madsen%
\ \protect \BOthers {.}}{%
{\protect \APACyear {2017}}%
}]{%
Madsen2017_BestPractices}
\APACinsertmetastar {%
Madsen2017_BestPractices}%
\begin{APACrefauthors}%
Madsen, A.%
, McKagan, S\BPBI B.%
\BCBL {}\ \BBA {} Sayre, E\BPBI C.%
\end{APACrefauthors}%
\unskip\
\newblock
\APACrefYearMonthDay{2017}{}{}.
\newblock
{\BBOQ}\APACrefatitle {Best Practices for Administering Concept Inventories}
  {Best practices for administering concept inventories}.{\BBCQ}
\newblock
\APACjournalVolNumPages{The Physics Teacher}{55}{9}{530-536}.
\newblock
\begin{APACrefURL} \url{https://doi.org/10.1119/1.5011826} \end{APACrefURL}
\newblock
\begin{APACrefDOI} \doi{10.1119/1.5011826} \end{APACrefDOI}
\PrintBackRefs{\CurrentBib}

\bibitem [\protect \citeauthoryear {%
Martinková%
\ \protect \BOthers {.}}{%
Martinková%
\ \protect \BOthers {.}}{%
{\protect \APACyear {2017}}%
}]{%
martinkova_checking_2017}
\APACinsertmetastar {%
martinkova_checking_2017}%
\begin{APACrefauthors}%
Martinková, P.%
, Drabinová, A.%
, Liaw, Y\BHBI L.%
, Sanders, E\BPBI A.%
, McFarland, J\BPBI L.%
\BCBL {}\ \BBA {} Price, R\BPBI M.%
\end{APACrefauthors}%
\unskip\
\newblock
\APACrefYearMonthDay{2017}{{\APACmonth{06}}}{}.
\newblock
{\BBOQ}\APACrefatitle {Checking {Equity}: {Why} {Differential} {Item}
  {Functioning} {Analysis} {Should} {Be} a {Routine} {Part} of {Developing}
  {Conceptual} {Assessments}} {Checking {Equity}: {Why} {Differential} {Item}
  {Functioning} {Analysis} {Should} {Be} a {Routine} {Part} of {Developing}
  {Conceptual} {Assessments}}.{\BBCQ}
\newblock
\APACjournalVolNumPages{LSE}{16}{2}{rm2}.
\newblock
\begin{APACrefURL} \url{https://www.lifescied.org/doi/10.1187/cbe.16-10-0307}
  \end{APACrefURL}
\newblock
\begin{APACrefDOI} \doi{10.1187/cbe.16-10-0307} \end{APACrefDOI}
\PrintBackRefs{\CurrentBib}

\bibitem [\protect \citeauthoryear {%
McDermott%
\ \BBA {} Redish%
}{%
McDermott%
\ \BBA {} Redish%
}{%
{\protect \APACyear {1999}}%
}]{%
McDermott1999}
\APACinsertmetastar {%
McDermott1999}%
\begin{APACrefauthors}%
McDermott, L\BPBI C.%
\BCBT {}\ \BBA {} Redish, E\BPBI F.%
\end{APACrefauthors}%
\unskip\
\newblock
\APACrefYearMonthDay{1999}{}{}.
\newblock
{\BBOQ}\APACrefatitle {Resource Letter: PER-1: Physics Education Research}
  {Resource letter: Per-1: Physics education research}.{\BBCQ}
\newblock
\APACjournalVolNumPages{American Journal of Physics}{67}{9}{755--767}.
\PrintBackRefs{\CurrentBib}

\bibitem [\protect \citeauthoryear {%
{National Research Council}%
}{%
{National Research Council}%
}{%
{\protect \APACyear {2012}}%
}]{%
NRC:2012}
\APACinsertmetastar {%
NRC:2012}%
\begin{APACrefauthors}%
{National Research Council}.%
\end{APACrefauthors}%
\unskip\
\newblock
\APACrefYear{2012}.
\newblock
\APACrefbtitle {Discipline-Based Education Research: Understanding and
  Improving Learning in Undergraduate Science and Engineering}
  {Discipline-based education research: Understanding and improving learning in
  undergraduate science and engineering}.
\newblock
\APACaddressPublisher{Washington, DC}{The National Academies Press}.
\PrintBackRefs{\CurrentBib}

\bibitem [\protect \citeauthoryear {%
Naylor%
}{%
Naylor%
}{%
{\protect \APACyear {2024}}%
}]{%
Naylor:2024fci}
\APACinsertmetastar {%
Naylor:2024fci}%
\begin{APACrefauthors}%
Naylor, W.%
\end{APACrefauthors}%
\unskip\
\newblock
\APACrefYearMonthDay{2024}{}{}.
\newblock
{\BBOQ}\APACrefatitle {The Multi-Million-Dollar Inventory: The {FCI} Uses and
  Applications} {The multi-million-dollar inventory: The {FCI} uses and
  applications}.{\BBCQ}
\newblock
\BIn{} \APACrefbtitle {Proceedings of the Australian Conference on Science and
  Mathematics Education} {Proceedings of the australian conference on science
  and mathematics education}\ (\BVOL~30).
\newblock
\begin{APACrefURL}
  \url{https://openjournals.library.sydney.edu.au/IISME/article/view/18763}
  \end{APACrefURL}
\PrintBackRefs{\CurrentBib}

\bibitem [\protect \citeauthoryear {%
Naylor%
, Rubab%
, Wang%
\BCBL {}\ \BBA {} Xuelan%
}{%
Naylor%
\ \protect \BOthers {.}}{%
{\protect \APACyear {2024}}%
}]{%
Naylor:2024}
\APACinsertmetastar {%
Naylor:2024}%
\begin{APACrefauthors}%
Naylor, W.%
, Rubab, S.%
, Wang, Y\BHBI G.%
\BCBL {}\ \BBA {} Xuelan, Q.%
\end{APACrefauthors}%
\unskip\
\newblock
\APACrefYearMonthDay{2024}{}{}.
\newblock
{\BBOQ}\APACrefatitle {Issues {Relating} to {CAT} and {Gain} in {Pyschometric}
  testing on {STEM} subjects using {IRT}} {Issues {Relating} to {CAT} and
  {Gain} in {Pyschometric} testing on {STEM} subjects using {IRT}}.{\BBCQ}
\newblock
\BIn{} \APACrefbtitle {Contemporary {Approaches} to {Research} in {STEM}
  {Education}} {Contemporary {Approaches} to {Research} in {STEM} {Education}}\
  (\BVOL~5).
\newblock
\APACaddressPublisher{}{Cambridge Scholars Press}.
\PrintBackRefs{\CurrentBib}

\bibitem [\protect \citeauthoryear {%
Nissen%
, Many~Horses%
, Dusen%
, Jariwala%
\BCBL {}\ \BBA {} Close%
}{%
Nissen%
\ \protect \BOthers {.}}{%
{\protect \APACyear {2022}}%
}]{%
nissen_providing_2022}
\APACinsertmetastar {%
nissen_providing_2022}%
\begin{APACrefauthors}%
Nissen, J\BPBI M.%
, Many~Horses, I\BPBI H.%
, Dusen, B\BPBI V.%
, Jariwala, M.%
\BCBL {}\ \BBA {} Close, E.%
\end{APACrefauthors}%
\unskip\
\newblock
\APACrefYearMonthDay{2022}{{\APACmonth{03}}}{}.
\newblock
{\BBOQ}\APACrefatitle {Providing {Context} for {Identifying} {Effective}
  {Introductory} {Mechanics} {Courses}} {Providing {Context} for {Identifying}
  {Effective} {Introductory} {Mechanics} {Courses}}.{\BBCQ}
\newblock
\APACjournalVolNumPages{The Physics Teacher}{60}{3}{179--182}.
\newblock
\begin{APACrefURL}
  [{2026-09-16}]\url{https://pubs.aip.org/pte/article/60/3/179/2848250/Providing-Context-for-Identifying-Effective}
  \end{APACrefURL}
\newblock
\begin{APACrefDOI} \doi{10.1119/5.0023763} \end{APACrefDOI}
\PrintBackRefs{\CurrentBib}

\bibitem [\protect \citeauthoryear {%
Nissen%
, Talbot%
, Nasim~Thompson%
\BCBL {}\ \BBA {} Van~Dusen%
}{%
Nissen%
\ \protect \BOthers {.}}{%
{\protect \APACyear {2018}}%
}]{%
Nissen:2018}
\APACinsertmetastar {%
Nissen:2018}%
\begin{APACrefauthors}%
Nissen, J\BPBI M.%
, Talbot, R\BPBI M.%
, Nasim~Thompson, A.%
\BCBL {}\ \BBA {} Van~Dusen, B.%
\end{APACrefauthors}%
\unskip\
\newblock
\APACrefYearMonthDay{2018}{Mar}{}.
\newblock
{\BBOQ}\APACrefatitle {Comparison of normalized gain and Cohen's $d$ for
  analyzing gains on concept inventories} {Comparison of normalized gain and
  cohen's $d$ for analyzing gains on concept inventories}.{\BBCQ}
\newblock
\APACjournalVolNumPages{Phys. Rev. Phys. Educ. Res.}{14}{}{010115}.
\newblock
\begin{APACrefURL}
  \url{https://link.aps.org/doi/10.1103/PhysRevPhysEducRes.14.010115}
  \end{APACrefURL}
\newblock
\begin{APACrefDOI} \doi{10.1103/PhysRevPhysEducRes.14.010115} \end{APACrefDOI}
\PrintBackRefs{\CurrentBib}

\bibitem [\protect \citeauthoryear {%
Qiu%
, Li%
\BCBL {}\ \BBA {} Wang%
}{%
Qiu%
\ \protect \BOthers {.}}{%
{\protect \APACyear {2019}}%
}]{%
Qiu:2019}
\APACinsertmetastar {%
Qiu:2019}%
\begin{APACrefauthors}%
Qiu, X\BHBI L.%
, Li, X.%
\BCBL {}\ \BBA {} Wang, W\BHBI C.%
\end{APACrefauthors}%
\unskip\
\newblock
\APACrefYearMonthDay{2019}{}{}.
\newblock
{\BBOQ}\APACrefatitle {Differential Item Functioning in Diagnostic
  Classification Models} {Differential item functioning in diagnostic
  classification models}.{\BBCQ}
\newblock
\BIn{} M.~von Davier\ \BBA {} Y\BHBI S.~Lee\ (\BEDS), \APACrefbtitle {Handbook
  of Diagnostic Classification Models: Models and Model Extensions,
  Applications, Software Packages} {Handbook of diagnostic classification
  models: Models and model extensions, applications, software packages}\
  (\BPGS\ 379--393).
\newblock
\APACaddressPublisher{Cham}{Springer International Publishing}.
\newblock
\begin{APACrefURL} \url{https://doi.org/10.1007/978-3-030-05584-4_18}
  \end{APACrefURL}
\newblock
\begin{APACrefDOI} \doi{10.1007/978-3-030-05584-4_18} \end{APACrefDOI}
\PrintBackRefs{\CurrentBib}

\bibitem [\protect \citeauthoryear {%
{R Core Team}%
}{%
{R Core Team}%
}{%
{\protect \APACyear {2024}}%
}]{%
R-base}
\APACinsertmetastar {%
R-base}%
\begin{APACrefauthors}%
{R Core Team}.%
\end{APACrefauthors}%
\unskip\
\newblock
\APACrefYearMonthDay{2024}{}{}.
\newblock
{\BBOQ}\APACrefatitle {R: A Language and Environment for Statistical Computing}
  {R: A language and environment for statistical computing}{\BBCQ}\
  [\bibcomputersoftwaremanual].
\newblock
\APACaddressPublisher{Vienna, Austria}{}.
\newblock
\begin{APACrefURL} \url{http://www.R-project.org/} \end{APACrefURL}
\PrintBackRefs{\CurrentBib}

\bibitem [\protect \citeauthoryear {%
Rasch%
}{%
Rasch%
}{%
{\protect \APACyear {1960}}%
}]{%
Rasch:1960}
\APACinsertmetastar {%
Rasch:1960}%
\begin{APACrefauthors}%
Rasch, G.%
\end{APACrefauthors}%
\unskip\
\newblock
\APACrefYear{1960}.
\newblock
\APACrefbtitle {Probabilistic models for some intelligence and attainment
  tests} {Probabilistic models for some intelligence and attainment tests}.
\newblock
\APACaddressPublisher{}{University of Chicago Press}.
\PrintBackRefs{\CurrentBib}

\bibitem [\protect \citeauthoryear {%
Redish%
}{%
Redish%
}{%
{\protect \APACyear {2003}}%
}]{%
Redish2003}
\APACinsertmetastar {%
Redish2003}%
\begin{APACrefauthors}%
Redish, E\BPBI F.%
\end{APACrefauthors}%
\unskip\
\newblock
\APACrefYear{2003}.
\newblock
\APACrefbtitle {Teaching Physics with the Physics Suite} {Teaching physics with
  the physics suite}.
\newblock
\APACaddressPublisher{}{Wiley}.
\PrintBackRefs{\CurrentBib}

\bibitem [\protect \citeauthoryear {%
Singer%
, Nielsen%
\BCBL {}\ \BBA {} Schweingruber%
}{%
Singer%
\ \protect \BOthers {.}}{%
{\protect \APACyear {2012}}%
}]{%
Singer2012}
\APACinsertmetastar {%
Singer2012}%
\begin{APACrefauthors}%
Singer, S\BPBI R.%
, Nielsen, N\BPBI R.%
\BCBL {}\ \BBA {} Schweingruber, H\BPBI A.%
\end{APACrefauthors}%
\unskip\
\newblock
\APACrefYear{2012}.
\newblock
\APACrefbtitle {DBER Report} {Dber report}.
\newblock
\APACaddressPublisher{}{National Academies Press}.
\PrintBackRefs{\CurrentBib}

\bibitem [\protect \citeauthoryear {%
Steyer%
}{%
Steyer%
}{%
{\protect \APACyear {2001}}%
}]{%
steyer_classical_2001}
\APACinsertmetastar {%
steyer_classical_2001}%
\begin{APACrefauthors}%
Steyer, R.%
\end{APACrefauthors}%
\unskip\
\newblock
\APACrefYearMonthDay{2001}{}{}.
\newblock
{\BBOQ}\APACrefatitle {Classical ({Psychometric}) {Test} {Theory}} {Classical
  ({Psychometric}) {Test} {Theory}}.{\BBCQ}
\newblock
\BIn{} \APACrefbtitle {International {Encyclopedia} of the {Social} \&
  {Behavioral} {Sciences}} {International {Encyclopedia} of the {Social} \&
  {Behavioral} {Sciences}}\ (\BPGS\ 1955--1962).
\newblock
\APACaddressPublisher{}{Elsevier}.
\newblock
\begin{APACrefURL}
  [{2024-02-07}]\url{https://linkinghub.elsevier.com/retrieve/pii/B008043076700721X}
  \end{APACrefURL}
\newblock
\begin{APACrefDOI} \doi{10.1016/B0-08-043076-7/00721-X} \end{APACrefDOI}
\PrintBackRefs{\CurrentBib}

\bibitem [\protect \citeauthoryear {%
Svetina%
\ \protect \BOthers {.}}{%
Svetina%
\ \protect \BOthers {.}}{%
{\protect \APACyear {2018}}%
}]{%
svetina_examining_2018}
\APACinsertmetastar {%
svetina_examining_2018}%
\begin{APACrefauthors}%
Svetina, D.%
, Feng, Y.%
, Paulsen, J.%
, Valdivia, M.%
, Valdivia, A.%
\BCBL {}\ \BBA {} Dai, S.%
\end{APACrefauthors}%
\unskip\
\newblock
\APACrefYearMonthDay{2018}{{\APACmonth{05}}}{}.
\newblock
{\BBOQ}\APACrefatitle {Examining {DIF} in the {Context} of {CDMs} {When} the
  {Q}-{Matrix} {Is} {Misspecified}} {Examining {DIF} in the {Context} of {CDMs}
  {When} the {Q}-{Matrix} {Is} {Misspecified}}.{\BBCQ}
\newblock
\APACjournalVolNumPages{Front. Psychol.}{9}{}{696}.
\newblock
\begin{APACrefURL}
  \url{http://journal.frontiersin.org/article/10.3389/fpsyg.2018.00696/full}
  \end{APACrefURL}
\newblock
\begin{APACrefDOI} \doi{10.3389/fpsyg.2018.00696} \end{APACrefDOI}
\PrintBackRefs{\CurrentBib}

\bibitem [\protect \citeauthoryear {%
Wang%
\ \BBA {} Bao%
}{%
Wang%
\ \BBA {} Bao%
}{%
{\protect \APACyear {2010}}%
}]{%
Wang:2010}
\APACinsertmetastar {%
Wang:2010}%
\begin{APACrefauthors}%
Wang, J.%
\BCBT {}\ \BBA {} Bao, L.%
\end{APACrefauthors}%
\unskip\
\newblock
\APACrefYearMonthDay{2010}{}{}.
\newblock
{\BBOQ}\APACrefatitle {Analyzing force concept inventory with item response
  theory} {Analyzing force concept inventory with item response theory}.{\BBCQ}
\newblock
\APACjournalVolNumPages{American Journal of Physics}{78}{10}{1064-1070}.
\newblock
\begin{APACrefURL} \url{https://doi.org/10.1119/1.3443565} \end{APACrefURL}
\newblock
\begin{APACrefDOI} \doi{10.1119/1.3443565} \end{APACrefDOI}
\PrintBackRefs{\CurrentBib}

\bibitem [\protect \citeauthoryear {%
Yasuda%
, Mae%
, Hull%
\BCBL {}\ \BBA {} Taniguchi%
}{%
Yasuda%
\ \protect \BOthers {.}}{%
{\protect \APACyear {2021}}%
}]{%
Yasuda:2021}
\APACinsertmetastar {%
Yasuda:2021}%
\begin{APACrefauthors}%
Yasuda, J.%
, Mae, N.%
, Hull, M\BPBI M.%
\BCBL {}\ \BBA {} Taniguchi, M.%
\end{APACrefauthors}%
\unskip\
\newblock
\APACrefYearMonthDay{2021}{May}{}.
\newblock
{\BBOQ}\APACrefatitle {Analysis to Develop Computerized Adaptive Testing with
  the Force Concept Inventory} {Analysis to develop computerized adaptive
  testing with the force concept inventory}.{\BBCQ}
\newblock
\APACjournalVolNumPages{Journal of Physics: Conference
  Series}{1929}{1}{012009}.
\newblock
\begin{APACrefURL} \url{https://doi.org/10.1088/1742-6596/1929/1/012009}
  \end{APACrefURL}
\newblock
\begin{APACrefDOI} \doi{10.1088/1742-6596/1929/1/012009} \end{APACrefDOI}
\PrintBackRefs{\CurrentBib}

\end{thebibliography}

\clearpage
\appendix


\section{The Q-Matrix}

The Q-matrix for this study is shown in Table \ref{tab:LeQ} below. 
It shows the our original baseline Q-matrix; the post-validated one using Qval(GDINA) in R \citep{R-base} and 
the one developed by \citet{le:2025}. Note that in most cases our original Q-matrix coding agrees 
with \citet{le:2025}, except for items: $19, 26$ \& $27$. 

\begin{table}[htbp]
\centering
\begin{threeparttable}
\caption{The Q-Matrix used for G-DINA models in this work}
\label{tab:LeQ}
\begin{tabular}{rccc c rccc c rccc}
\toprule
& \multicolumn{3}{c}{Items 1--10} & & & \multicolumn{3}{c}{Items 11--20} & & & \multicolumn{3}{c}{Items 21--30} \\
\cmidrule(lr){2-4} \cmidrule(lr){7-9} \cmidrule(lr){12-14}
Item & ${}^\dag$Orig. & ${}^\ddag$Sug. & ${}^*$Le & & Item & ${}^\dag$Orig. & ${}^\ddag$Sug. & ${}^*$Le  & & Item & ${}^\dag$Orig. & ${}^\ddag$Sug. & ${}^*$Le  \\
\midrule
1 & \multicolumn{3}{c}{010} & \quad & 11 & 100 & 1\underline{1}\underline{1} & 100 & \quad & 21 & 100 & 1\underline{1}\underline{1} & 100 \\
2 & 010 & \underline{1}1\underline{1} & 010 & \quad & 12 & \multicolumn{3}{c}{110} & \quad & 22 & 100 & 1\underline{1}\underline{1} & 100 \\
3 & 010 & \underline{1}10 & 010 & \quad & 13 & \multicolumn{3}{c}{100} & \quad & 23 & \multicolumn{3}{c}{110} \\
4 & 010 & \underline{1}10 & 010 & \quad & 14 & \multicolumn{3}{c}{110} & \quad & 24 & 010 & \underline{1}1\underline{1} & 010 \\
5 & \multicolumn{3}{c}{100} & \quad & 15 & 010 & \underline{1}1\underline{1} & 010 & \quad & 25 & 110 & 11\underline{1} & 110 \\
6 & \multicolumn{3}{c}{010} & \quad & 16 & \multicolumn{3}{c}{010} & \quad & 26 & 110 & 11\underline{1} & 100 \\
7 & \multicolumn{3}{c}{110} & \quad & 17 & 100 & 1\underline{1}\underline{1} & 100 & \quad & 27 & \multicolumn{2}{c}{110} & 100 \\
8 & 100 & 1\underline{1}0 & 100 & \quad & 18 & \multicolumn{3}{c}{100} & \quad & 28 & \multicolumn{3}{c}{100} \\
9 & 100 & 1\underline{1}0 & 100 & \quad & 19 & \multicolumn{2}{c}{011} & 010 & \quad & 29 & \multicolumn{3}{c}{100} \\
10 & 010 & \underline{1}1\underline{1} & 010 & \quad & 20 & 011 & \underline{1}11 & 011 & \quad & 30 & 100 & 1\underline{1}\underline{1} & 100 \\
\bottomrule
\end{tabular}
\vskip 0.2cm
\begin{tablenotes}
\small
\item ${}^\dag$Original baseline Q-matrix; ${}^\ddag$Suggested post-validated Q-matrix; ${}^*$\citet{le:2025} Q-matrix.
Underlined entries indicate modifications suggested by the empirical validation algorithm: \textit{Qval} \citep[][]{R-base} 
used in our analyses. Entries centered across more than one column indicates identical specifications across models, where
each entry corresponds to the three skills: $k_1,k_2,k_3$ in Table \ref{table:skills}. 
\end{tablenotes}
\end{threeparttable}
\end{table}

\section{Guessing and Slipping Parameters}

Here we present some details of the DIF analyses for the slip and guessing parameters compared between both groups: 
the US and UJ cohorts. The table can be used in conjunction with Figure \ref{fig:fci_dif_plot} and  
Table \ref{tab:fci_dif_cohen_h}.


\begin{table}[ht]
\centering
\caption{Supplementary Item Parameter Estimates (Guessing and Slipping) Across UJ and LASSO Cohorts} 
\label{tab:supp_item_params}
\begin{tabular}{lrrrr}
  \hline
Item & $g_{UJ}$ & $s_{UJ}$ & $g_{LASSO}$ & $s_{LASSO}$ \\ 
  \hline
   1 & 0.390 & 0.223 & 0.526 & 0.080 \\ 
   2 & 0.192 & 0.833 & 0.244 & 0.448 \\ 
   3 & 0.408 & 0.802 & 0.362 & 0.268 \\ 
   4 & 0.478 & 0.196 & 0.165 & 0.394 \\ 
   5 & 0.090 & 0.681 & 0.087 & 0.401 \\ 
   6 & 0.378 & 0.381 & 0.617 & 0.062 \\ 
   7 & 0.285 & 0.433 & 0.489 & 0.089 \\ 
   8 & 0.256 & 0.432 & 0.480 & 0.111 \\ 
   9 & 0.214 & 0.594 & 0.076 & 0.973 \\ 
   10 & 0.117 & 0.879 & 0.309 & 0.123 \\ 
   11 & 0.143 & 0.756 & 0.088 & 0.278 \\ 
   12 & 0.337 & 0.290 & 0.545 & 0.074 \\ 
   13 & 0.151 & 0.384 & 0.088 & 0.126 \\ 
   14 & 0.213 & 0.429 & 0.251 & 0.197 \\ 
   15 & 0.111 & 0.964 & 0.164 & 0.497 \\ 
   16 & 0.112 & 0.965 & 0.399 & 0.139 \\ 
   17 & 0.577 & 0.501 & 0.078 & 0.558 \\ 
   18 & 0.194 & 0.377 & 0.093 & 0.273 \\ 
   19 & 0.277 & 0.000 & 0.079 & 0.992 \\ 
   20 & 0.313 & 0.742 & 0.044 & 0.000 \\ 
   21 & 0.137 & 0.871 & 0.130 & 0.564 \\ 
   22 & 0.120 & 0.881 & 0.349 & 0.237 \\ 
   23 & 0.070 & 0.959 & 0.234 & 0.933 \\ 
   24 & 0.084 & 0.927 & 0.465 & 0.083 \\ 
   25 & 0.103 & 0.924 & 0.131 & 0.430 \\ 
   26 & 0.171 & 0.843 & 0.035 & 0.557 \\ 
   27 & 0.240 & 0.824 & 0.443 & 0.125 \\ 
   28 & 0.363 & 0.163 & 0.216 & 0.145 \\ 
   29 & 0.496 & 0.184 & 0.381 & 0.226 \\ 
   30 & 0.036 & 0.930 & 0.118 & 0.252 \\ 
   \hline
\end{tabular}
\end{table}


As can be seen for Item 19 in Table \ref{tab:supp_item_params} near-zero response variance 
in one subgroup during optimization causes parameter covariance matrix 
estimation to degenerate, driving the Wald denominator toward zero. The magnitude of this test statistic reflects
numerical instability in matrix inversion rather than extreme substantive DIF; item magnitude can 
instead be evaluated using the non-parametric Cohen's h effect size.

\section{DIF Comparisons}

This section notes a comparison of Differential Item Functioning (DIF) via Wald and LR methods using a validated Q-matrix 
for our original Q-matrix in Table \ref{tab:LeQ}. Upon running the function $QVal()$ in R we found the following item 
comparisons.

\begin{table}[ht!]
\centering
\caption{Comparison of Differential Item Functioning (DIF) via Wald ($W_{stat}$) and LR Methods using validated Q-matrices 
 } 
\label{tab:step2_dif_comparison}
\begin{threeparttable}
\begin{tabular}{lrrrr}
  \hline
  Item   & $W_{stat}$ & $^\dagger p^W_{adj}$ & $LR_{stat}$ & $^\dagger p^{LR}_{adj}$ \\ 
  \hline
   1 & 29.64 & 0.00 & 23.43 & 0.00 \\ 
   2 & 104.85 & 0.00 & 58.61 & 0.00 \\ 
   3 & 139.26 & 0.00 & 75.56 & 0.00 \\ 
   4 & 100.26 & 0.00 & 75.10 & 0.00 \\ 
   5 & 182.80 & 0.00 &  &  \\ 
   6 & 146.03 & 0.00 & 135.68 & 0.00 \\ 
   7 & 96.70 & 0.00 & 80.62 & 0.00 \\ 
   8 & 115.93 & 0.00 & 109.05 & 0.00 \\ 
   9 & 150.07 & 0.00 & 171.03 & 0.00 \\ 
   10 & 827.60 & 0.00 & 250.19 & 0.00 \\ 
   11 & 196.90 & 0.00 & 120.00 & 0.00 \\ 
   12 & 71.22 & 0.00 & 61.39 & 0.00 \\ 
   13 & 181.78 & 0.00 &  &  \\ 
   14 & 2.62 & 0.62 & 5.47 & 0.24 \\ 
   15 & 126.78 & 0.00 & 78.94 & 0.00 \\ 
   16 & 1318.02 & 0.00 & 445.44 & 0.00 \\ 
   17 & 549.19 & 0.00 &  &  \\ 
   18 & 124.15 & 0.00 &  &  \\ 
   19 & 185.57 & 0.00 & 279.95 & 0.00 \\ 
   20 & 180.35 & 0.00 & 47.05 & 0.00 \\ 
   21 & 44.66 & 0.00 & 34.79 & 0.00 \\ 
   22 & 421.79 & 0.00 & 178.33 & 0.00 \\ 
   23 & 186.63 & 0.00 & 114.05 & 0.00 \\ 
   24 & 1381.48 & 0.00 & 616.74 & 0.00 \\ 
   25 & 21.30 & 0.00 & 24.24 & 0.00 \\ 
   26 & 65.17 & 0.00 & 102.81 & 0.00 \\ 
   27 & 244.89 & 0.00 & 191.96 & 0.00 \\ 
   28 & 65.48 & 0.00 & 52.62 & 0.00 \\ 
   29 & 32.73 & 0.00 & 34.59 & 0.00 \\ 
   30 & 107.33 & 0.00 & 85.38 & 0.00 \\ 
   \hline
\end{tabular}
\vskip 0.2cm
\begin{tablenotes}
	\footnotesize{
	\item[$\dagger$] Adjusted $p$-values are presented for each method.
    }
    \end{tablenotes}
     \end{threeparttable}
\end{table}

\section{Comparison of Cohen's d and h}

The table below shows comparisons of the effect sizes using Cohen's $d$ or $h$ \citep{cohen_statistical_2013}, where 
Cohen's $d$ over-estimates item DIFs.

\begin{table}[ht]
\centering
\caption{FCI Item-Level DIF Analysis between LASSO ($N = 4,750$) and UJ ($N = 1,016$) Cohorts$^\dagger$} 
\label{tab:dif-cohen-comp}
\begin{threeparttable}
\rotatebox{90}{%
\begin{tabular}{rrrrrrlrrrl}
  \hline
Item & $P_{LASSO}$ & $P_{UJ}$ & $d$ & $CI_d$ & $DIF_d$ & $h$ & $CI_h$ & $DIF_h$ \\ 
  \hline
    1 & 0.670 & 0.470 &  0.412 & [ 0.344,  0.481] & Moderate (*) &  0.407 & [ 0.339,  0.475] & Moderate (*) \\ 
    2 & 0.350 & 0.190 &  0.366 & [ 0.298,  0.434] & Moderate (*) &  0.364 & [ 0.296,  0.432] & Moderate (*) \\ 
    3 & 0.490 & 0.370 &  0.244 & [ 0.176,  0.312] & Moderate (*) &  0.243 & [ 0.175,  0.311] & Moderate (*) \\ 
    4 & 0.320 & 0.540 & -0.456 & [-0.524, -0.388] & Moderate (*) & -0.448 & [-0.516, -0.381] & Moderate (*) \\ 
    5 & 0.170 & 0.140 &  0.083 & [ 0.015,  0.151] & Negligible   &  0.083 & [ 0.015,  0.151] & Negligible   \\ 
    6 & 0.730 & 0.430 &  0.638 & [ 0.569,  0.707] & High (**)    &  0.618 & [ 0.551,  0.686] & High (**)    \\ 
    7 & 0.620 & 0.340 &  0.584 & [ 0.515,  0.652] & High (**)    &  0.568 & [ 0.500,  0.636] & High (**)    \\ 
    8 & 0.600 & 0.310 &  0.609 & [ 0.540,  0.677] & High (**)    &  0.591 & [ 0.523,  0.659] & High (**)    \\ 
    9 & 0.070 & 0.260 & -0.530 & [-0.598, -0.461] & High (**)    & -0.535 & [-0.602, -0.467] & High (**)    \\ 
   10 & 0.510 & 0.120 &  0.925 & [ 0.855,  0.995] & High (**)    &  0.883 & [ 0.816,  0.951] & High (**)    \\ 
   11 & 0.200 & 0.170 &  0.077 & [ 0.010,  0.145] & Negligible   &  0.077 & [ 0.010,  0.145] & Negligible   \\ 
   12 & 0.670 & 0.420 &  0.519 & [ 0.450,  0.587] & High (**)    &  0.508 & [ 0.440,  0.575] & High (**)    \\ 
   13 & 0.220 & 0.260 & -0.094 & [-0.162, -0.026] & Negligible   & -0.094 & [-0.161, -0.026] & Negligible   \\ 
   14 & 0.420 & 0.280 &  0.297 & [ 0.229,  0.365] & Moderate (*) &  0.295 & [ 0.227,  0.363] & Moderate (*) \\ 
   15 & 0.290 & 0.100 &  0.494 & [ 0.426,  0.562] & Moderate (*) &  0.494 & [ 0.426,  0.562] & Moderate (*) \\ 
   16 & 0.560 & 0.100 &  1.122 & [ 1.051,  1.192] & High (**)    &  1.048 & [ 0.980,  1.115] & High (**)    \\ 
   17 & 0.140 & 0.560 & -0.981 & [-1.051, -0.911] & High (**)    & -0.924 & [-0.992, -0.856] & High (**)    \\ 
   18 & 0.200 & 0.290 & -0.210 & [-0.278, -0.143] & Moderate (*) & -0.210 & [-0.278, -0.142] & Moderate (*) \\ 
   19 & 0.060 & 0.280 & -0.613 & [-0.681, -0.544] & High (**)    & -0.620 & [-0.688, -0.553] & High (**)    \\ 
   20 & 0.390 & 0.280 &  0.235 & [ 0.167,  0.303] & Moderate (*) &  0.234 & [ 0.166,  0.302] & Moderate (*) \\ 
   21 & 0.180 & 0.140 &  0.109 & [ 0.041,  0.177] & Negligible   &  0.109 & [ 0.042,  0.177] & Negligible   \\ 
   22 & 0.420 & 0.120 &  0.718 & [ 0.649,  0.787] & High (**)    &  0.703 & [ 0.635,  0.770] & High (**)    \\ 
   23 & 0.200 & 0.060 &  0.426 & [ 0.357,  0.494] & Moderate (*) &  0.432 & [ 0.365,  0.500] & Moderate (*) \\ 
   24 & 0.630 & 0.080 &  1.404 & [ 1.332,  1.477] & High (**)    &  1.260 & [ 1.193,  1.328] & High (**)    \\ 
   25 & 0.200 & 0.090 &  0.316 & [ 0.248,  0.384] & Moderate (*) &  0.318 & [ 0.250,  0.386] & Moderate (*) \\ 
   26 & 0.100 & 0.170 & -0.206 & [-0.274, -0.138] & Moderate (*) & -0.206 & [-0.274, -0.139] & Moderate (*) \\ 
   27 & 0.570 & 0.210 &  0.794 & [ 0.725,  0.863] & High (**)    &  0.759 & [ 0.691,  0.827] & High (**)    \\ 
   28 & 0.330 & 0.470 & -0.289 & [-0.357, -0.221] & Moderate (*) & -0.287 & [-0.355, -0.219] & Moderate (*) \\ 
   29 & 0.450 & 0.570 & -0.242 & [-0.310, -0.174] & Moderate (*) & -0.241 & [-0.308, -0.173] & Moderate (*) \\ 
   30 & 0.230 & 0.040 &  0.579 & [ 0.510,  0.647] & High (**)    &  0.598 & [ 0.530,  0.665] & High (**)    \\ 
   \hline
\end{tabular}%
}
\vskip 0.2cm
\begin{tablenotes}
    \footnotesize
    \item[$\dagger$] $P_{\text{LASSO}}$ and $P_{\text{UJ}}$ denote the success rates for each cohort, respectively.
    \item[*] Positive values of $d$ and $h$ indicate a performance advantage for the LASSO cohort ($P_{\text{LASSO}} > P_{\text{UJ}}$); negative values indicate an advantage for the UJ cohort ($P_{\text{UJ}} > P_{\text{LASSO}}$). Item 15 ($h = 0.494$) falls under the $0.500$ threshold in three-decimal reporting.
\end{tablenotes}
\end{threeparttable}
\end{table}

\end{document}